\documentclass{jfm}

\usepackage{amsmath,bm}
\usepackage{placeins}
\hypersetup{
  unicode=false,
  pdfencoding=pdfdoc,
  bookmarksnumbered=true,
  bookmarksopen=true,
  pdftitle={Transient and asymptotic Taylor--Aris dispersion of Brownian rods in arbitrary regular-polygonal ducts},
  pdfauthor={Jingsen Feng and Xu Chu}
}
\graphicspath{{fig/fig1/}{fig/fig2/}{fig/fig3/}{fig/fig4/}{fig/fig5/}{fig/fig6/}{fig/fig7/}{fig/fig8/}{fig/fig10/}{fig/fig11/}{fig/fig12/}{fig/fig13/}{fig/fig14/}}

\newcommand{\Dbar}{\bar D}
\newcommand{\Pe}{Pe}
\newcommand{\Per}{Pe_r}
\newcommand{\dd}{\,\mathrm d}
\newcommand{\avg}[1]{\left\langle #1\right\rangle}
\newcommand{\x}{\bm x}
\newcommand{\pvec}{\bm p}
\newcommand{\e}{\bm e}

\lefttitle{J. Feng and X. Chu}
\righttitle{Brownian rods in regular-polygon channels}

\title{Transient and asymptotic Taylor--Aris dispersion of Brownian rods in arbitrary regular-polygonal ducts}

\author{Jingsen Feng\aff{1} \and Xu Chu\aff{1}}

\affiliation{\aff{1}Department of Engineering, University of Exeter, Exeter EX4 4QF, United Kingdom}

\corresau{Xu Chu, \email{x.chu@exeter.ac.uk}}

\makeatletter
\def\jfm@simpleoddfoot{\hbox to \textwidth{\hfill{\cppagefont\thepage}}}
\def\jfm@simpleevenfoot{\hbox to \textwidth{{\cppagefont\thepage}\hfill}}
\def\oddabsfooterflag{\jfm@simpleoddfoot}
\def\evenabsfooterflag{\jfm@simpleevenfoot}
\def\pagelimitfooter{\jfm@simpleevenfoot}
\def\ps@titlepage{\leftskip\z@\let\@mkboth\@gobbletwo\vfuzz=5\p@
  \def\@oddhead{\vbox{\vspace*{-4pt}\hbox to \textwidth{\@j@urnal\hfill}}}%
  \def\@evenhead{\vbox{\vspace*{-4pt}\hbox to \textwidth{\@j@urnal\hfill}}}%
  \def\@oddfoot{\jfm@simpleoddfoot}%
  \def\@evenfoot{\jfm@simpleevenfoot}%
  \def\sectionmark##1{}%
  \def\subsectionmark##1{}%
}
\def\ps@headings{\let\@mkboth\markboth
  \def\@oddhead{\hfill{\itshape\@righttitle}\hfill}%
  \def\@evenhead{\hfill\itshape\@lefttitle\hfill}%
  \def\@oddfoot{\jfm@simpleoddfoot}%
  \def\@evenfoot{\jfm@simpleevenfoot}%
  \def\sectionmark##1{\markboth{##1}{}}%
  \def\subsectionmark##1{\markright{##1}}%
}
\makeatother
\begin{document}
\maketitle

\begin{abstract}
Taylor--Aris dispersion of Brownian rods in non-circular ducts is governed by a coupling absent from passive-scalar theory. Pressure-driven shear aligns the rods and makes translational diffusion tensorial, while duct geometry determines how this tensor is sampled across the cross-section. We formulate this problem for dilute rods in regular-polygonal ducts of arbitrary side number. At each cross-sectional point, a local shear-aligned Jeffery--Brownian closure gives four transport fields, namely two transverse diffusivities, a direct axial diffusivity and a signed shear--axial cross coefficient. Because the shear frame rotates through a polygon, these fields enter a conservative two-dimensional transverse operator rather than a radial scalar-diffusion problem. Its zero mode is a non-uniform invariant density, which replaces the area measure in the Taylor--Aris reduction and reduces, in the circular-pipe limit, to a weighting proportional to the inverse shear-direction diffusivity.

The resulting cell problem separates the effects of rod alignment on streamline sampling and transverse relaxation. Alignment produces only a small, non-monotone shift in mean speed, but gives a larger enhancement of the Taylor coefficient by reducing transverse mixing. Normalization by the same-geometry spherical coefficient removes most passive shape dependence and exposes the approach to the fully aligned transverse-mixing limit. Finite regular polygons converge smoothly to the circular-pipe branch, whereas low-sided polygons retain distinct shear-sampling signatures. A biorthogonal spectral formulation resolves finite-time releases. Localized, multi-peaked and broad injections excite different non-zero transverse modes and exhibit different pre-asymptotic variance growth, but modal decay selects the common long-time Taylor--Aris coefficient given by the cell problem.
\end{abstract}

\section{Introduction}
\label{sec:introduction}

Taylor--Aris dispersion describes the axial spreading produced when transverse diffusion samples a shear flow. In a circular tube, a scalar solute crosses the parabolic velocity profile by radial diffusion, and the long-time concentration evolves as a one-dimensional cloud with mean speed equal to the area-averaged flow and an enhanced axial diffusivity \citep{taylor1953,aris1956}. Subsequent work has extended this picture to finite-time spreading, non-circular conduits, shaped channels, boundary effects, pulsatile or actively controlled walls, colloidal transport and microfluidic settings \citep{guell1987,vedel2014,aminian2015,aminian2016,marbach2019,salerno2020,lee2021,alessio2022,chang2023,guan2024}. Generalized Taylor dispersion supplies the corresponding cell-problem formulation for more complex transport operators and heterogeneous diffusivities \citep{ramirez2006,alexandre2021}. Here we use this viewpoint in a setting where the transverse mixing problem is no longer scalar or radially organized. For Brownian rods in a non-circular pressure-driven duct, particle orientation makes the translational diffusivity tensorial, while the cross-section determines the two-dimensional shear field over which that tensor acts.

Non-circular ducts alter Taylor dispersion even for a passive scalar, because the Poiseuille velocity and the transverse relaxation modes are geometry-dependent rather than radial \citep{guell1987,aminian2015,aminian2016,lee2021,chang2023}. Rectangular, triangular, hexagonal and other polygonal channels are common idealizations in micro- and minichannel flow models \citep{tamayol2010}, while idealized pore geometries are also used to describe dispersion in porous materials \citep{liu2024compaction,liu2024scaling,liu2026mechanism}. Their fully developed pressure-driven flow is governed by a two-dimensional Poisson/torsion-analogy problem, and the wall geometry determines both the velocity contours and the shear distribution \citep{shah1975,shah1978,tamayol2010}. Classical duct-flow and heat-transfer studies have treated arbitrary and regular-polygonal cross-sections as canonical departures from the circular tube \citep{cheng1967,cheng1969,shah1975,shah1978}. More recent analytical and semi-analytical descriptions of laminar flow in non-circular microchannels have again used regular polygons as a useful family connecting the equilateral triangle, square, higher-sided ducts and the circular limit \citep{tamayol2010}. Regular polygons are a natural choice here because the same calculation has to work for finite-\(N\) sections, where no radial reduction is available, and for the circular limit, where such a reduction must be recovered.

The rotation of an ellipsoid in a linear Stokes flow follows Jeffery dynamics \citep{jeffery1922}, while Brownian rotary diffusion turns Jeffery's orbit family into a shear-dependent probability density on orientation space \citep{leal1971,hinch1972,hinch1973,brenner1974}. In simple shear, slender particles spend longer near streamwise alignment as the rotational P{\'e}clet number increases \citep{hinch1972,stover1992,leahy2015}; related orientational transport has been measured and modelled through shear-enhanced rotational diffusion \citep{leahy2013,leahy2015,peng2024} and through the dynamics of individual Brownian rods in microchannel flow \citep{zottl2019}. Pressure-driven flows add spatial variation to this local orientation bias and can produce cross-stream migration through shear-dependent orientation statistics and finite-length or non-local effects \citep{nitsche1997,schiek1997}.

The translational Brownian motion of a rod is anisotropic even at zero shear: the diffusivity parallel to the particle axis differs from that in the two transverse directions, as follows from the distinct parallel and perpendicular resistance functions set by the aspect ratio \citep{perrin1936,tirado1979,tirado1984,han2006,kraft2013}. Single-particle experiments on ellipsoidal and other shaped colloids have made the coupling between particle geometry, translational diffusion and rotational diffusion directly observable \citep{han2006,han2009,chakrabarty2013,kraft2013}. A shear-biased orientation distribution therefore makes the orientation-averaged translational diffusivity a tensor in the laboratory frame, consistent with generalized Taylor-dispersion treatments of orientable Brownian particles in homogeneous shear \citep{frankel1993}. In a locally simple shear, the tensor contains two transverse components, an axial component and a signed shear--axial cross coefficient. A scalar diffusivity cannot retain the directional roles needed in a Taylor--Aris reduction.

Orientation-controlled dispersion has also been identified in active and complex suspensions. Experiments and simulations of swimming cells in shear and Poiseuille flows show that orientation can suppress cross-stream motility, change drift and concentrate particles in particular shear regions \citep{zottl2012,zottl2013,rusconi2014,croze2013}. Continuum theories for pressure-driven active suspensions, gyrotactic pipe flow and active Brownian particles have connected such non-uniform cross-sectional distributions to longitudinal dispersivity, including the effects of particle shape, wall accumulation, upstream swimming and finite-time relaxation \citep{ezhilan2015,chilukuri2015,jiang2019,jiang2020,peng2020,wang2021,jiang2021}. For elongated microswimmers in pressure-driven channels, aspect-ratio-dependent shear trapping and centreline collapse have been predicted \citep{vennamneni2020migration}, and the associated longitudinal dispersion can exhibit anomalous scaling \citep{vennamneni2025anomalous}. Related active-particle Taylor--Aris studies have treated pre-asymptotic focusing, anisotropic diffusion, buoyancy--flow coupling, diffuse reflection and oscillatory forcing \citep{guan2023preasymptotic,guan2024anisodiff,wang2025spheroid,wang2025,zeng2025diffuse}. Although those problems include swimming and, in some cases, orientational drift or boundary accumulation, they support the broader point that shear-biased orientation and non-uniform cross-sectional equilibria can control confined transport. The present work instead considers passive dilute rods in polygonal pressure-driven flow, for which Jeffery--Brownian orientation statistics and anisotropic translational diffusion are the only particle-level mechanisms retained.

Passive Brownian rods have been analysed most directly in planar channel geometry. In the theory and Monte Carlo calculations of \citet{kumar2021}, the cross-stream coordinate is unique and the rod orientation is described by one in-plane angle. Jeffery alignment then appears as a shear-dependent reduction of cross-stream diffusivity and produces a larger longitudinal Taylor coefficient than for a sphere with the same orientationally averaged diffusivity. \citet{khair2022} derived the corresponding small- and large-\(\Per\) asymptotic limits for the mean speed and dispersivity. These studies identify the rod-specific alignment mechanism, but their transverse mixing problem remains one-dimensional. A circular tube already requires radial shear variation and a full three-dimensional orientation distribution, as treated in the circular-tube counterpart of the present tensorial rod theory \citep{feng2026tube}. In a regular-polygonal duct, the geometry removes even the radial organization: both the magnitude and the direction of the Poiseuille gradient vary over the section, so the shear-aligned diffusivity tensor must be placed and differentiated in a genuinely two-dimensional cross-section.

The absence of a global radial coordinate is therefore the organizing geometric difficulty of the polygonal problem. At each cross-sectional point, the Poiseuille gradient defines a local shear plane and a down-gradient direction. The Jeffery--Brownian angular problem is still local and geometry-independent once the local shear strength and aspect ratio are specified \citep{jeffery1922,leal1971,hinch1972,brenner1974}. The polygonal geometry determines where the resulting coefficients are placed and how the physical divergence acts on their fluxes. In the local shear frame, the orientation-averaged translational diffusivity separates into diffusion along the down-gradient shear direction, diffusion along the transverse direction perpendicular to that shear plane, diffusion along the duct axis, and a signed coupling between the shear direction and the duct axis. We denote these four scalar coefficients by \(D_s\), \(D_\eta\), \(B\) and \(A\), respectively. Because the local shear frame rotates through the polygon, the conservative transverse operator is not obtained by simply replacing the radial derivative in the circular-tube theory. It is a genuinely two-dimensional operator involving the coefficient fields and the shear direction \(\e_s(\x)\), as in non-circular duct problems where the cross-sectional geometry must be retained explicitly \citep{shah1975,shah1978,tamayol2010}.

The leading cross-sectional state is therefore not the area measure. Writing \(\rho_\infty\) for the normalized invariant density of the circular branch, the no-flux condition reduces to \(\mathrm d(D_s\rho_\infty)/\mathrm dr=0\), giving \(\rho_\infty(r)\propto D_s^{-1}(r)\), as in the circular-tube rod problem \citep{feng2026tube}. The relaxed rod cloud therefore gives greater weight to regions where shear alignment has reduced radial mobility. In finite regular polygons the same mechanism gives a two-dimensional invariant density \(\rho_N(\x)\). This density sets the sampled mean velocity. The same transverse relaxation operator gives the Taylor cell problem, while \(B\) contributes direct axial diffusion and \(A\) gives a lower-order conservative drift correction. This component-wise use of the transport operator follows the generalized Taylor-dispersion viewpoint for non-uniform and anisotropic transport \citep{ramirez2006,alexandre2021,guan2024anisodiff}. The finite-time problem then asks how an injection that is not initially proportional to \(\rho_N\) relaxes through the non-zero transverse modes before its axial variance reaches the Taylor--Aris regime \citep{vedel2012,vedel2014,jiang2021,jiang2026oscillatory}.

We therefore formulate a Taylor--Aris theory for dilute Brownian rods in pressure-driven flow through regular-polygonal ducts. The local steady orientation Fokker--Planck equation is solved once as a function of shear strength and aspect ratio, and its second moments provide \(D_s\), \(D_\eta\), \(B\) and \(A\). These coefficients are mapped onto the polygonal Poiseuille field using a local shear-aligned frame. A conservative cross-sectional transport equation then yields the invariant density, the leading sampling speed, the Taylor cell problem, the direct axial diffusivity and the cross-diffusive drift. The finite-polygon coefficients are compared with the circular-pipe branch as \(N\to\infty\), separating the rod-induced alignment effect from the passive geometric dependence of the Poiseuille cell problem.

The transient part of the theory uses the same transverse relaxation operator before the cross-section has equilibrated. We construct a biorthogonal spectral model whose zero mode is the invariant density and whose non-zero modes carry the memory of the injection profile. This gives a reduced description of localized, multi-peaked and broad initial distributions, resolves how their cross-sectional memory decays, and predicts the corresponding time-dependent axial variance. The finite-time variance growth then shows how injection-dependent pre-asymptotic spreading crosses over to the long-time Taylor--Aris coefficient.

The argument below follows this sequence. Section~\ref{sec:geometry-local-closure} defines the polygonal geometry, local shear coordinates and Jeffery--Brownian closure. Section~\ref{sec:cross-sectional-transport} maps the local tensorial coefficients into the conservative transport equation and identifies the invariant cross-sectional density. Section~\ref{sec:taylor-aris-reduction} derives the one-dimensional Taylor--Aris reduction, including the distinct roles of \(D_s\), \(D_\eta\), \(A\) and \(B\). Section~\ref{sec:steady-effective-transport-results} reports the steady effective coefficients and the polygon-to-pipe convergence. Section~\ref{sec:transient-spectral-validation} uses the transverse spectrum to describe finite-time relaxation from different injections and to verify convergence of the transient axial variance-growth rate to the cell-problem Taylor coefficient.

\section{Geometry, shear coordinates and local rod closure}
\label{sec:geometry-local-closure}

For a fully developed pressure-driven flow in a regular-polygonal duct, the velocity is axial but its transverse gradient is not organized by a single radial coordinate, except in the circular limit. A Brownian rod at a given cross-sectional point therefore experiences a locally simple shear whose shear plane is determined by \(\nabla_\perp u_N\), where \(\nabla_\perp\) denotes the gradient with respect to the cross-sectional coordinates. This section defines that local shear frame and uses it to convert the Jeffery--Brownian orientational equilibrium into dimensionless translational transport coefficients. The construction assumes dilute point rods, local affine shear, and no explicit wall-induced orientational potential; wall and corner effects enter through the Poiseuille shear field. Under these assumptions, the angular problem is geometry-independent once the local rotational P{\'e}clet number \(q_N(\x;\Per)\) and the rod aspect ratio \(p\) are specified, while the polygonal geometry determines where those local coefficients are placed in the cross-section.

\subsection{Polygonal channel and Poiseuille flow}
\label{subsec:polygon-poiseuille}

Let \(R_{\rm in}\) be the inradius of the dimensional polygonal cross-section, and write \(\x=\x^\ast/R_{\rm in}\) for the dimensionless transverse coordinate. The scaled channel is the infinite prism
\begin{equation}
  \mathcal D_N=\Omega_N\times\mathbb R_z ,
\end{equation}
where \(\Omega_N\) is a regular \(N\)-sided polygon with unit inradius. For finite \(N\),
\begin{equation}
  \Omega_N=
  \left\{\x\in\mathbb R^2:
  \bm n_j\cdot\x<1,\quad j=0,\ldots,N-1
  \right\},
\label{eq:polygon-domain}
\end{equation}
with outward unit normals
\begin{equation}
  \bm n_j=
  \left(
  \cos\frac{2\pi j}{N},
  \sin\frac{2\pi j}{N}
  \right).
\label{eq:polygon-normals}
\end{equation}
The area is
\begin{equation}
  |\Omega_N|=N\tan\frac{\pi}{N}.
\label{eq:polygon-area}
\end{equation}
The notation \(N=\infty\) denotes the circular limit, \(\Omega_\infty=\{\x:|\x|<1\}\).

For each cross-section the dimensional pressure-driven velocity is proportional to the solution of the standard duct-flow torsion problem \citep{shah1975,shah1978,tamayol2010}
\begin{equation}
  -\Delta_\perp \tilde u_N=1,
  \qquad
  \tilde u_N=0\quad\hbox{on }\partial\Omega_N .
\label{eq:polygon-torsion}
\end{equation}
We use the centreline normalization
\begin{equation}
  u_N(\x)=
  \frac{\tilde u_N(\x)}
       {\max_{\Omega_N}\tilde u_N},
\label{eq:velocity-normalization}
\end{equation}
so that the axial flow is \(\bm u_f=Uu_N(\x)\e_z\) before non-dimensionalization, and \(u_N\e_z\) after scaling the velocity by \(U\). In the circular limit this convention gives
\begin{equation}
  u_\infty(r)=1-r^2 .
\label{eq:circular-profile}
\end{equation}
The maximum-speed normalization separates the shape of the Poiseuille field from the overall speed \(U\). The comparisons below fix \(\Per\), the maximum local rotational P{\'e}clet number, or equivalently the maximum shear rate relative to rotational diffusion.

\subsection{Shear-coordinate convention}
\label{subsec:shear-coordinate}

Unlike the circular tube, a regular polygon has no global radial coordinate aligned with the Poiseuille gradient throughout the section. The direction of \(\nabla_\perp u_N\) changes with position and is distorted by flat sides and corners. A local shear coordinate is therefore required before the rod closure can be applied. We use a shear-aligned frame \((\e_s,\e_\eta,\e_z)\) tied to the local Poiseuille gradient. The signed cross-diffusion coefficient uses the orientation of this frame; throughout the paper the \(s\)-axis points down the Poiseuille gradient according to
\begin{equation}
  \e_s(\x)=
  -\frac{\nabla_\perp u_N(\x)}
        {|\nabla_\perp u_N(\x)|}
\label{eq:es-definition}
\end{equation}
wherever \(|\nabla_\perp u_N|>0\). Thus \(\e_s\) points from faster streamlines towards slower streamlines. In the circular limit,
\begin{equation}
  \nabla_\perp u_\infty=-2r\e_r,
  \qquad
  \e_s=\e_r ,
\label{eq:circular-es}
\end{equation}
which makes the polygon notation reduce directly to the usual radial coordinate in a tube. The transverse direction completing the local shear-plane basis is
\begin{equation}
  \e_\eta=\bm R_{\pi/2}\e_s ,
\label{eq:eeta-definition}
\end{equation}
where \(\bm R_{\pi/2}\) denotes a counter-clockwise rotation in the cross-sectional plane. At isolated points where the shear vanishes, \(\e_s\) may be chosen arbitrarily; the zero-shear closure below gives \(D_s=D_\eta\) and \(A=0\).

With this convention, the dimensionless velocity gradient has only one transverse derivative in the local frame:
\begin{equation}
  \partial_s u_N=-|\nabla_\perp u_N|,
  \qquad
  \partial_\eta u_N=0 .
\label{eq:local-simple-shear}
\end{equation}
Thus the orientation dynamics at a fixed cross-sectional point reduce to those in a locally simple shear in the \(s\)-\(z\) plane. Consequently, the angular closure depends on the polygon only through the scalar shear strength \(q_N(\x)\).

The dimensional local shear rate is
\begin{equation}
  \dot\gamma_N(\x)
  =
  \frac{U}{R_{\rm in}}|\nabla_\perp u_N(\x)| .
\label{eq:local-shear-rate}
\end{equation}
The scalar \(q_N\) is the local rotational P{\'e}clet number used in the angular Smoluchowski problem. With the convention in Appendix~\ref{app:local-orientation-solver},
\begin{equation}
  q_N(\x)=\frac{\dot\gamma_N(\x)}{2D_\theta},
  \qquad
  \Per=\frac{\dot\gamma_{\max}}{2D_\theta},
\label{eq:rotational-peclet-physical}
\end{equation}
where \(D_\theta\) is the rotational diffusivity and \(\dot\gamma_{\max}=\max_{\Omega_N}\dot\gamma_N\). Hence
\begin{equation}
  q_N(\x;\Per)
  =
  \Per\,
  \frac{|\nabla_\perp u_N(\x)|}
       {\max_{\Omega_N}|\nabla_\perp u_N|}
  .
\label{eq:qN-definition}
\end{equation}
With this choice, \(\max_{\Omega_N}q_N=\Per\). It therefore has the same interpretation for a triangle, a square, a many-sided polygon and the circular reference case. A circular-compatible normalization based on \(q_\infty=\Per r\) is useful for limiting checks, but the results in the main geometry comparisons use \eqref{eq:qN-definition}.

\begin{figure}
  \centering
  \includegraphics[width=0.98\linewidth]{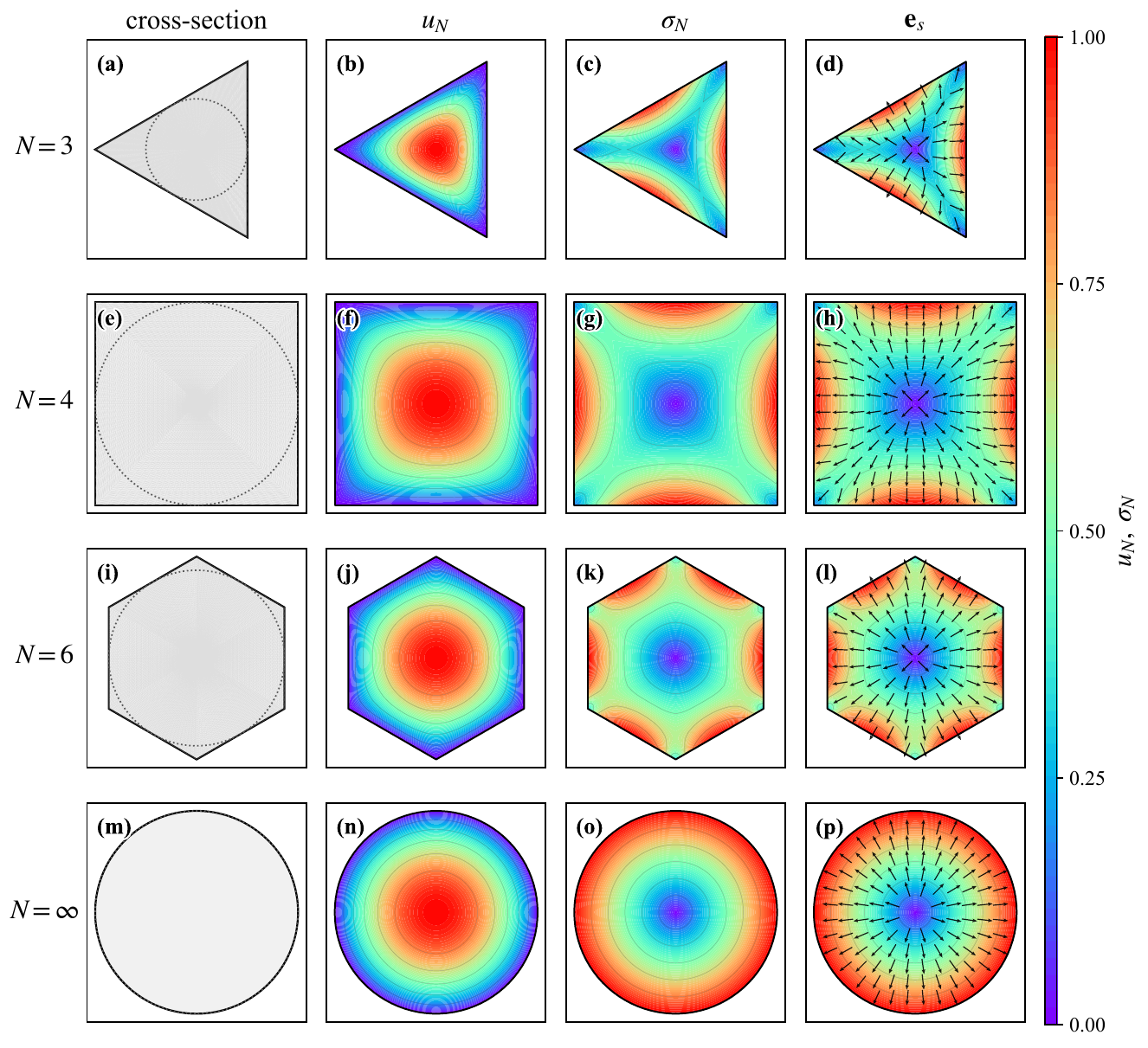}
  \caption{Geometry-to-shear map for representative cross-sections. The rows correspond to \(N=3\), \(4\), \(6\) and \(\infty\). The columns show the polygonal domain \(\Omega_N\), the normalized Poiseuille velocity \(u_N\), the normalized shear magnitude \(\sigma_N=|\nabla_\perp u_N|/\max_{\Omega_N}|\nabla_\perp u_N|\), and the down-gradient shear direction \(\e_s\). The circular row gives \(u_\infty=1-r^2\) and \(\e_s=\e_r\).}
  \label{fig:geometry-shear}
\end{figure}

The shear maps in Figure~\ref{fig:geometry-shear} make the departure from the circular pipe explicit. At finite \(N\), the largest shear forms side-wall bands, while the circular limit organizes the shear on radial shells. Near the polygon corners, the shear-plane direction is also distorted. These features set both the locations of strongest Jeffery--Brownian alignment and the orientation of the signed cross coefficient \(A\). As \(N\) increases, the finite-polygon structure is smoothed and the radial pipe limit is recovered. The vector field in the last column records the sign convention in \eqref{eq:es-definition}; it is this directed field, not only the scalar shear magnitude, that fixes the sign of the later cross-diffusive flux.

\subsection{Local Jeffery--Brownian orientation closure}
\label{subsec:local-closure}

At each \(\x\), the rod orientation is described by a local equilibrium distribution obtained from the steady rotational Smoluchowski balance between Jeffery drift and rotational Brownian diffusion \citep{jeffery1922,leal1971,hinch1972,brenner1974}. The closure treats the rods as dilute point particles whose length is small compared with \(R_{\rm in}\) and with the local length over which the shear varies. Steric and hydrodynamic interactions with the wall are therefore neglected; these mechanisms can produce migration in related channel-flow theories \citep{nitsche1997,schiek1997}. Under this approximation the wall and corner effects enter only through the Poiseuille shear field, not through an explicit wall-dependent orientational potential.

We write the local distribution as
\begin{equation}
  g=g(\pvec;q,p),
  \qquad
  \pvec\in S^2 ,
\label{eq:orientation-density}
\end{equation}
where \(p\) is the particle aspect ratio and \(q\) is the local shear strength. The Jeffery shape factor is
\begin{equation}
  \beta(p)=\frac{p^2-1}{p^2+1}.
\label{eq:beta-definition}
\end{equation}
In invariant form the angular balance is
\begin{equation}
  \nabla_p\cdot
  \left[
  \nabla_p g-\dot{\pvec}_J(\pvec;q,p)g
  \right]=0,
  \qquad
  \int_{S^2}g\,\dd\Omega=1 ,
\label{eq:main-angular-smoluchowski}
\end{equation}
where \(\nabla_p\) is the surface gradient on \(S^2\) and \(\dot{\pvec}_J\) is the dimensionless Jeffery drift in the local \(s\)-\(z\) shear. Appendix~\ref{app:local-orientation-solver} gives the coordinate form and discretization. For \(p=1\), \(\beta=0\) and the steady distribution is isotropic. For \(p>1\), Jeffery rotation is anisotropic: rods rotate more slowly near streamwise orientations than near cross-stream orientations \citep{hinch1972,stover1992,leahy2015}. Rotational Brownian diffusion regularizes this residence-time bias and produces a steady, shear-dependent orientational distribution. The angular problem is solved once as a function of \((q,p)\) and then reused for every polygonal cross-section.

Let
\begin{equation}
  p_s=\pvec\cdot\e_s,\qquad
  p_\eta=\pvec\cdot\e_\eta,\qquad
  p_z=\pvec\cdot\e_z ,
\end{equation}
and denote local orientational averages by
\begin{equation}
  \avg{f}_q=\int_{S^2} f(\pvec)g(\pvec;q,p)\,\dd\Omega .
\label{eq:orientation-average}
\end{equation}
The particle-level diffusivities are specified by Perrin's expressions for a prolate spheroid \citep{perrin1934,perrin1936}. Let \(a_p\) be the semi-major axis, \(\eta_f\) the dynamic viscosity of the suspending fluid, \(k_{\rm B}\) Boltzmann's constant and \(T\) the absolute temperature. The translational diffusivity along the rod axis and that in either perpendicular direction are
\begin{subequations}
\label{eq:perrin-translational-diffusivities}
\begin{align}
  D_\parallel
  &=
  \frac{k_{\rm B}T}{16\pi\eta_f a_p}\,p
  \left[
  -\frac{2p}{p^2-1}
  +
  \frac{2p^2-1}{(p^2-1)^{3/2}}
  \log\!\left(
  \frac{p+\sqrt{p^2-1}}{p-\sqrt{p^2-1}}
  \right)
  \right],
  \label{eq:perrin-Dparallel}\\
  D_\perp
  &=
  \frac{k_{\rm B}T}{16\pi\eta_f a_p}\,p
  \left[
  \frac{p}{p^2-1}
  +
  \frac{2p^2-3}{(p^2-1)^{3/2}}
  \log\!\left(p+\sqrt{p^2-1}\right)
  \right].
  \label{eq:perrin-Dperp}
\end{align}
\end{subequations}
The corresponding rotational diffusivity used in \eqref{eq:rotational-peclet-physical} is
\begin{equation}
  D_\theta
  =
  \frac{3k_{\rm B}T}{16\pi\eta_f a_p^3}
  \frac{p^4}{p^4-1}
  \left[
  \frac{(2p^2-1)\log\!\left(p+\sqrt{p^2-1}\right)}
       {p\sqrt{p^2-1}}
  -1
  \right].
\label{eq:perrin-rotational-diffusivity}
\end{equation}
We normalize \(D_\parallel\) and \(D_\perp\) by the three-dimensional mean
\begin{equation}
  \Dbar=\frac{D_\parallel+2D_\perp}{3},
\end{equation}
and write
\begin{equation}
  d_\parallel=\frac{D_\parallel}{\Dbar},
  \qquad
  d_\perp=\frac{D_\perp}{\Dbar},
  \qquad
  \frac{d_\parallel+2d_\perp}{3}=1 .
\label{eq:diffusivity-ratios}
\end{equation}
The spherical branch is obtained by taking the limiting isotropic value, giving \(d_\parallel=d_\perp=1\) for \(p=1\). In the infinitely slender limit, \(d_\parallel\to3/2\) and \(d_\perp\to3/4\).

The orientation-averaged translational diffusivity tensor, used in orientable-particle dispersion theory \citep{frankel1993}, is
\begin{equation}
  \bm D(q;p)
  =
  d_\perp\bm I
  +
  (d_\parallel-d_\perp)\avg{\pvec\pvec}_q .
\label{eq:orientation-averaged-tensor}
\end{equation}
The local transport functions in the shear basis are the projections
\begin{subequations}
\label{eq:local-functions}
\begin{align}
  D_s(q;p)
  &=
  d_\perp+
  (d_\parallel-d_\perp)
  \avg{p_s^2}_q,\\
  D_\eta(q;p)
  &=
  d_\perp+
  (d_\parallel-d_\perp)
  \avg{p_\eta^2}_q,\\
  B(q;p)
  &=
  d_\perp+
  (d_\parallel-d_\perp)
  \avg{p_z^2}_q,\\
  A(q;p)
  &=
  (d_\parallel-d_\perp)
  \avg{p_sp_z}_q .
\end{align}
\end{subequations}
Here \(D_s\) is the diffusivity in the down-gradient shear direction, \(D_\eta\) is the transverse diffusivity perpendicular to the local shear plane, \(B\) is the direct axial diffusivity, and \(A\) is the signed shear-plane cross coefficient. Reversing the convention for \(\e_s\) would leave \(D_s\), \(D_\eta\) and \(B\) unchanged but would reverse \(A\), so \eqref{eq:es-definition} fixes the sign of all cross-diffusive fluxes.

The diagonal coefficients determine how rapidly a rod cloud relaxes across streamlines and spreads along the duct. The off-diagonal coefficient \(A\) is different: it is non-zero only when the orientational distribution is tilted in the shear plane, and it changes sign with the chosen \(s\)-direction. It is therefore a signed measure of shear-induced coupling between transverse and axial gradients, rather than a scalar enhancement of diffusion.

The trace normalization gives
\begin{equation}
  D_s+D_\eta+B=3,
\label{eq:trace-closure}
\end{equation}
and the diffusion tensor in the local \(s\)-\(z\) plane is positive definite:
\begin{equation}
  D_s>0,
  \qquad
  D_sB-A^2>0 .
\label{eq:closure-positive}
\end{equation}
Alignment increases the streamwise orientational moment \(\avg{p_z^2}_q\) and therefore enhances \(B\). The same redistribution reduces the transverse moments \(\avg{p_s^2}_q\) and \(\avg{p_\eta^2}_q\), weakening diffusion across and out of the local shear plane. The angular density and second moments behind this redistribution are shown for a representative slender-rod case in Figure~\ref{fig:app-orientation-density-moments} of Appendix~\ref{app:local-orientation-solver}.

In the spherical limit,
\begin{equation}
  p=1
  \quad\Rightarrow\quad
  D_s=D_\eta=B=1,\qquad A=0,
\label{eq:sphere-limit}
\end{equation}
for every value of \(q\).

\begin{figure}
  \centering
  \includegraphics[width=0.96\linewidth]{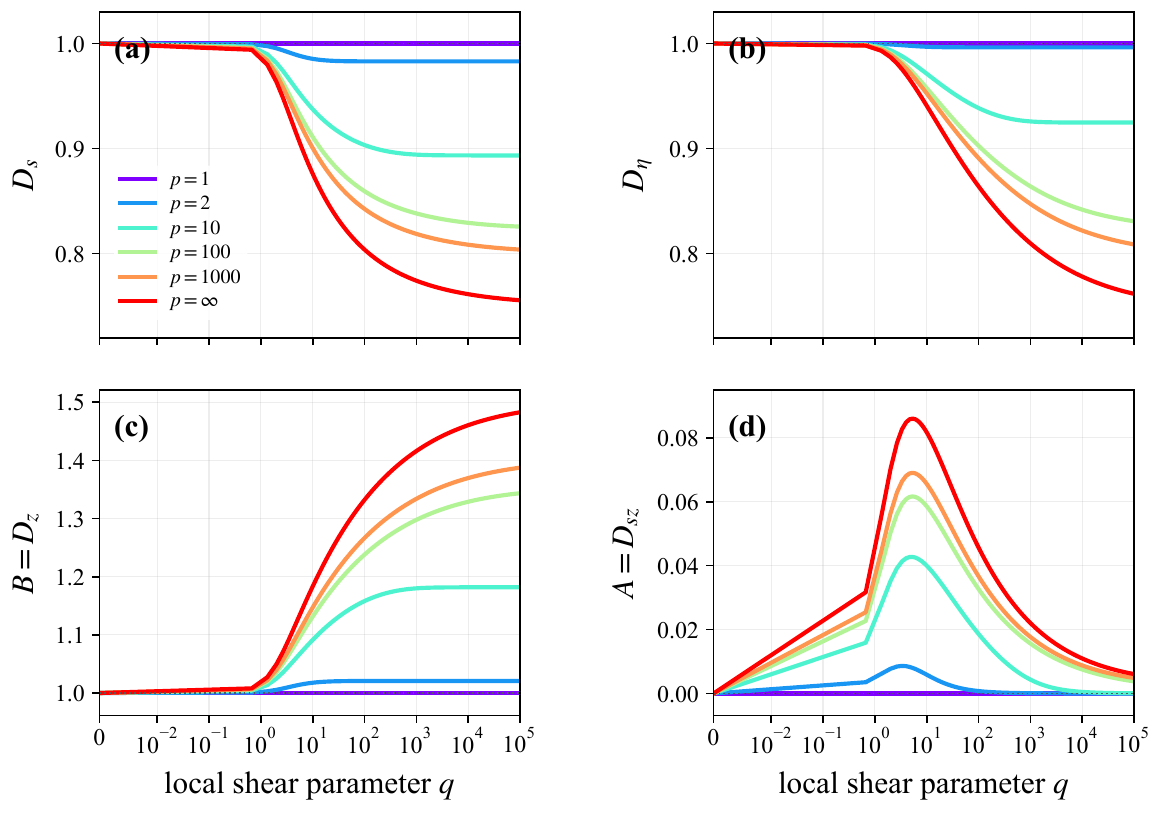}
  \caption{Local Jeffery--Brownian transport closure. The panels show \(D_s(q;p)\), \(D_\eta(q;p)\), \(B(q;p)\) and \(A(q;p)\) as functions of the local shear strength \(q\) for different aspect ratios \(p\). Spheres give the constant isotropic branch \(D_s=D_\eta=B=1\) and \(A=0\). Increasing \(p\) strengthens the shear-induced redistribution of translational diffusivity.}
  \label{fig:local-closure}
\end{figure}

Figure~\ref{fig:local-closure} gives the geometry-independent local response used for every polygon. The diagonal coefficients show the main effect of local alignment: transverse mixing is weakened, while the axial component is enhanced. The non-monotonicity of \(A\) follows directly from the mixed moment \(\avg{p_sp_z}_q\). At small \(q\), the distribution is nearly isotropic and the signed moment cancels. At intermediate \(q\), the distribution is both tilted and sufficiently broad in the \(s\)-direction, giving the largest cross coefficient. At very large \(q\), the rods are almost streamwise, so \(p_s\) becomes small and the mixed moment decreases. Combining Figure~\ref{fig:geometry-shear} with Figure~\ref{fig:local-closure} gives the local fields used by the cross-sectional transport theory:
\begin{equation}
  \x
  \longmapsto
  q_N(\x;\Per)
  \longmapsto
  \{D_s,D_\eta,B,A\} .
\label{eq:geometry-to-closure}
\end{equation}
In the following section these scalar functions are rotated back into the polygonal cross-section: \(D_s\) and \(D_\eta\) form the transverse diffusion tensor, \(A\e_s\) gives the cross-diffusion vector, and the conservative flux form determines the invariant density sampled by Taylor--Aris dispersion.

\section{Cross-sectional transport equation and invariant density}
\label{sec:cross-sectional-transport}

The local closure of Section~\ref{sec:geometry-local-closure} specifies how a rod diffuses at a point where the shear strength and shear plane are known. In a polygonal duct, however, both quantities vary across the cross-section. The rod cloud therefore does not sample streamlines according to area measure alone: shear-induced alignment changes the transverse relaxation operator and selects a modified invariant measure. This section converts the local Jeffery--Brownian coefficients into a conservative cross-sectional transport equation and identifies the invariant density that weights the leading Taylor--Aris state.

\subsection{Spatial coefficient fields in the polygonal section}
\label{subsec:field-mapping}

For a fixed triple \((N,p,\Per)\), the shear map \eqref{eq:qN-definition} defines four cross-sectional scalar fields:
\begin{subequations}
\label{eq:polygon-fields}
\begin{align}
  D_s(\x)&=D_s(q_N(\x;\Per);p),\\
  D_\eta(\x)&=D_\eta(q_N(\x;\Per);p),\\
  B(\x)&=B(q_N(\x;\Per);p),\\
  A(\x)&=A(q_N(\x;\Per);p).
\end{align}
\end{subequations}
After this substitution, \(D_s\), \(D_\eta\), \(B\) and \(A\) are fields on \(\Omega_N\); an explicit \(q\)-argument will be shown only when the scalar closure functions are meant. The transverse part of the orientation-averaged diffusivity is collected in the tensor
\begin{equation}
  \bm K_N(\x)
  =
  D_s(\x)\e_s(\x)\e_s(\x)^T
  +
  D_\eta(\x)\e_\eta(\x)\e_\eta(\x)^T .
\label{eq:transverse-tensor}
\end{equation}
The shear-plane coupling with axial gradients is collected in the vector
\begin{equation}
  \bm a_N(\x)=A(\x)\e_s(\x).
\label{eq:cross-vector}
\end{equation}
The tensor \(\bm K_N\) characterizes the local transverse mobility after rotation into the physical cross-section. The corresponding relaxation operator is the conservative product form defined below. The vector \(\bm a_N\) is the transverse projection of the axial--shear cross diffusivity; it determines how an axial concentration gradient drives a transverse flux and, reciprocally, how transverse variations generate an axial flux. The dyadic construction makes \(\bm K_N\) insensitive to the sign of \(\e_s\), whereas \(\bm a_N\) changes sign with the shear-coordinate convention fixed in \eqref{eq:es-definition}. The circular limit gives \(\bm a_\infty=A(r)\e_r\), matching the radial--axial coefficient used in the tube reduction.

\subsection{Conservative projected transport equation}
\label{subsec:conservative-equation}

The conservative product derivatives in the reduced flux follow from averaging the translational Smoluchowski flux after the position--orientation density has been projected onto the local angular equilibrium, as in orientable-particle dispersion treatments \citep{frankel1993}. Let \(c(\x,z,t)\) denote the orientation-averaged number concentration in the long channel, write \(c_z=\partial_z c\), and define the directional derivatives
\begin{equation}
  \partial_s=\e_s\cdot\nabla_\perp,
  \qquad
  \partial_\eta=\e_\eta\cdot\nabla_\perp .
\label{eq:s-eta-derivatives}
\end{equation}
For clarity, write the local equilibrium in the physical cross-section as
\begin{equation*}
  \begin{aligned}
    g_N(\x,\pvec)
    &=
    g\!\left(p_s(\x),p_\eta(\x),p_z;q_N(\x),p\right),\\
    p_s(\x)&=\pvec\cdot\e_s(\x),
    \qquad
    p_\eta(\x)=\pvec\cdot\e_\eta(\x),
    \qquad
    p_z=\pvec\cdot\e_z .
  \end{aligned}
\end{equation*}
The local-equilibrium ansatz for the orientation-resolved density is
\begin{equation}
  n(\x,z,\pvec,t)=c(\x,z,t)g_N(\x,\pvec)
\label{eq:polygon-local-equilibrium-ansatz}
\end{equation}
The dimensionless translational diffusivity of a rod with orientation \(\pvec\) is
\begin{equation}
  \bm D_{\rm tr}(\pvec)
  =
  d_\perp\bm I
  +
  (d_\parallel-d_\perp)\pvec\pvec .
\label{eq:orientation-resolved-diffusivity}
\end{equation}
After inserting this ansatz, the orientation-averaged advective--diffusive flux can be written as
\begin{equation}
  \bm J
  =
  \Pe\,u_N c\,\e_z
  -
  \int_{S^2}
  \bm D_{\rm tr}(\pvec)
  \nabla[c\,g_N(\x,\pvec)]
  \,\dd\Omega ,
\label{eq:averaged-smoluchowski-flux}
\end{equation}
where \(\Pe=UR_{\rm in}/\Dbar\) is the axial P{\'e}clet number and \(\nabla=\nabla_\perp+\e_z\partial_z\). Because \(g_N\) depends on \(\x\) through both \(q_N(\x)\) and the local shear frame, the derivative acts on the local equilibrium as well as on \(c\). The local simple-shear distribution is symmetric under \(p_\eta\mapsto -p_\eta\), so the only non-zero off-diagonal component of the orientation-averaged diffusivity is the \(s\)-\(z\) component \(A\). The local-frame flux components are therefore
\begin{align*}
  J_s&=-\partial_s(D_s c)-A c_z,
  \qquad
  J_\eta=-\partial_\eta(D_\eta c),\\
  J_z&=\Pe\,u_N c-\partial_s(Ac)-B c_z .
\end{align*}
Rotating \(J_s\) and \(J_\eta\) back into the physical cross-section gives the transverse flux below. The closure is local-affine in the shear frame: wall and corner geometry enter through the physical divergence of the resulting flux and through the Poiseuille shear map, not through an additional wall-induced orientational potential.

After local equilibration in orientation, the spatial conservation law is written as
\begin{equation}
  c_t+\nabla_\perp\cdot\bm J_\perp+\partial_zJ_z=0,
\label{eq:polygon-conservation}
\end{equation}
where the transverse flux is
\begin{equation}
  \bm J_\perp
  =
  -\e_s\,\partial_s(D_s c)
  -\e_\eta\,\partial_\eta(D_\eta c)
  -\e_s A c_z ,
\label{eq:polygon-transverse-flux}
\end{equation}
and the axial flux is
\begin{equation}
  J_z
  =
  \Pe\,u_N c
  -
  \partial_s(Ac)
  -
  B c_z .
\label{eq:polygon-axial-flux}
\end{equation}
The derivatives in \eqref{eq:polygon-transverse-flux} act on the scalar products \(D_s c\) and \(D_\eta c\) along the local shear directions; the subsequent divergence in \eqref{eq:polygon-conservation} is taken in the physical cross-section. This distinction matters because the shear basis is not a global curvilinear coordinate system in a polygon. Equation~\eqref{eq:polygon-transverse-flux} is therefore the physical flux vector obtained after the local shear-frame projection, not a shorthand for \(-\bm K_N\nabla_\perp c\). Similarly, the term \(\partial_s(Ac)\) in \eqref{eq:polygon-axial-flux} is the directional derivative of the scalar \(Ac\) along \(\e_s\). It is distinct from the cross-sectional divergence \(\nabla_\perp\cdot(Ac\,\e_s)\), which contains the additional geometry term \(Ac\,\nabla_\perp\cdot\e_s\).

The no-penetration condition is imposed on the physical boundary of the polygon:
\begin{equation}
  \bm n\cdot\bm J_\perp=0,
  \qquad
  \x\in\partial\Omega_N .
\label{eq:polygon-wall-bc}
\end{equation}
The axial coordinate is unbounded in the asymptotic problem and is taken to be periodic only in direct numerical computations.

The structure of \eqref{eq:polygon-transverse-flux}--\eqref{eq:polygon-axial-flux} separates the roles of the four transport fields. The pair \(D_s,D_\eta\) controls transverse relaxation and therefore sets the sampling of streamlines. The coefficient \(B\) gives direct axial diffusion. The coefficient \(A\) appears twice in the same orientation-averaged flux: \(-\e_s A c_z\) is a transverse flux driven by an axial concentration gradient, and \(-\partial_s(Ac)\) is the axial flux generated by transverse variation of the signed shear-plane moment.

Equations~\eqref{eq:polygon-conservation}--\eqref{eq:polygon-wall-bc} are the conservative flux form obtained after averaging the translational Smoluchowski flux over the local orientation equilibrium. For constant scalar diffusion this reduces to the familiar Fickian structure. When rod alignment varies across the channel, the same averaging changes the leading transverse equilibrium.
The non-uniform equilibrium produced below is a consequence of the projection onto an \(\x\)-dependent angular equilibrium. A constant orientation-integrated amplitude \(c\) does not correspond to a uniform density in the full position--orientation phase space, because the angular distribution itself varies with the local shear. The reduced transverse operator therefore relaxes to its invariant measure \(\rho_N\), rather than to the area measure.

\subsection{Invariant cross-sectional density}
\label{subsec:invariant-density}

The long-time Taylor--Aris reduction begins with the transverse relaxation problem obtained by setting axial gradients to zero \citep{taylor1953,aris1956,ramirez2006,alexandre2021}. The corresponding operator is
\begin{equation}
  \mathcal L_{0,N}c
  =
  \nabla_\perp\cdot
  \left[
  \e_s\,\partial_s(D_s c)
  +
  \e_\eta\,\partial_\eta(D_\eta c)
  \right],
\label{eq:L0N}
\end{equation}
with the no-flux boundary condition obtained from \eqref{eq:polygon-transverse-flux} at \(c_z=0\). The invariant density \(\rho_N\) is defined by
\begin{subequations}
\label{eq:rhoN-definition}
\begin{align}
  \mathcal L_{0,N}\rho_N&=0,
  \qquad
  \int_{\Omega_N}\rho_N\,\dd A=1 ,
  \label{eq:rhoN-interior}\\
  \bm n\cdot
  \left[
  \e_s\,\partial_s(D_s\rho_N)
  +
  \e_\eta\,\partial_\eta(D_\eta\rho_N)
  \right]&=0,
  \qquad
  \x\in\partial\Omega_N .
  \label{eq:rhoN-boundary}
\end{align}
\end{subequations}
For spherical particles the coefficients are \(D_s=D_\eta=1\), and the normalized solution is the uniform density \(1/|\Omega_N|\). For rods in shear, the conservative derivatives in \eqref{eq:L0N} generally produce
\begin{equation}
  \rho_N(\x)\ne\frac{1}{|\Omega_N|}.
\label{eq:rho-nonuniform}
\end{equation}
The circular limit makes the mechanism explicit. When \(N=\infty\), \(\e_s=\e_r\), \(\e_\eta=\e_\phi\), and the invariant density is axisymmetric, \eqref{eq:rhoN-definition} reduces to
\begin{equation}
  \frac1r\frac{\dd}{\dd r}
  \left[
  r\frac{\dd}{\dd r}\left(D_s(r)\rho_\infty(r)\right)
  \right]=0,
  \qquad
  \rho_\infty(r)\propto D_s(r)^{-1}.
\label{eq:circular-rho-check}
\end{equation}
Regions in which alignment suppresses radial mobility are therefore sampled more strongly by the reduced invariant measure. This is not a thermodynamic accumulation caused by an imposed potential, but the no-flux state of the coarse-grained local-equilibrium projection. The same principle carries over to polygons, although the varying shear direction and the distinction between \(D_s\) and \(D_\eta\) make the invariant density genuinely two-dimensional. Thus the polygonal density \(\rho_N\,\dd A\) reduces to the weighted radial sampling measure \(rD_s^{-1}(r)\,\dd r\) used in the circular-tube reduction.

The density \(\rho_N\) is the first quantity through which local alignment affects the macroscopic transport. Anticipating the one-dimensional reduction, let \(C(z,t)\) denote the cross-sectional mass per unit axial length. The leading cross-sectional concentration is then
\begin{equation}
  c_0(\x,z,t)=\rho_N(\x)C(z,t),
\label{eq:c0-rho}
\end{equation}
and therefore determines the velocity sampled at leading order:
\begin{equation}
  \bar u_N=\int_{\Omega_N}u_N(\x)\rho_N(\x)\,\dd A .
\label{eq:ubar-rho-preview}
\end{equation}
Section~\ref{sec:taylor-aris-reduction} derives this relation from the averaged axial flux; here it is introduced only to emphasize the physical role of \(\rho_N\). The same operator \(\mathcal L_{0,N}\) also supplies the cell problem for the Taylor coefficient in the following section. Thus the field maps shown below determine both the invariant measure and the Taylor cell problem.

\subsection{Cross-sectional tensor fields}
\label{subsec:tensor-field-figures}

\begin{figure}
  \centering
  \includegraphics[width=0.98\linewidth]{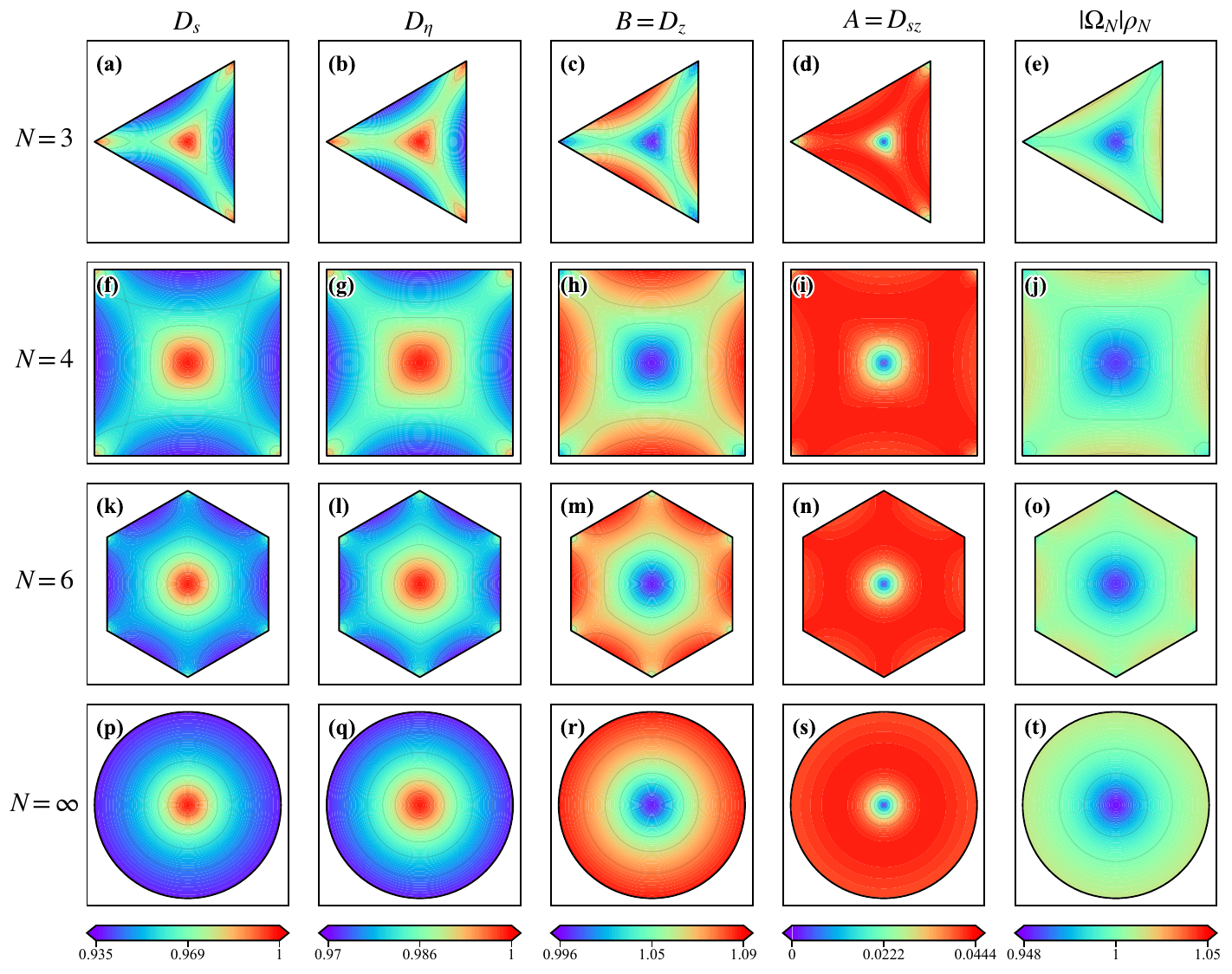}
  \caption{Cross-sectional transport fields for moderate rods and moderate orientational shear, \(p=10\) and \(\Per=10\). The rows correspond to \(N=3\), \(4\), \(6\) and \(\infty\). The columns show \(D_s\), \(D_\eta\), \(B\), \(A\) and \(|\Omega_N|\rho_N\).}
  \label{fig:fields-p10}
\end{figure}

\begin{figure}
  \centering
  \includegraphics[width=0.98\linewidth]{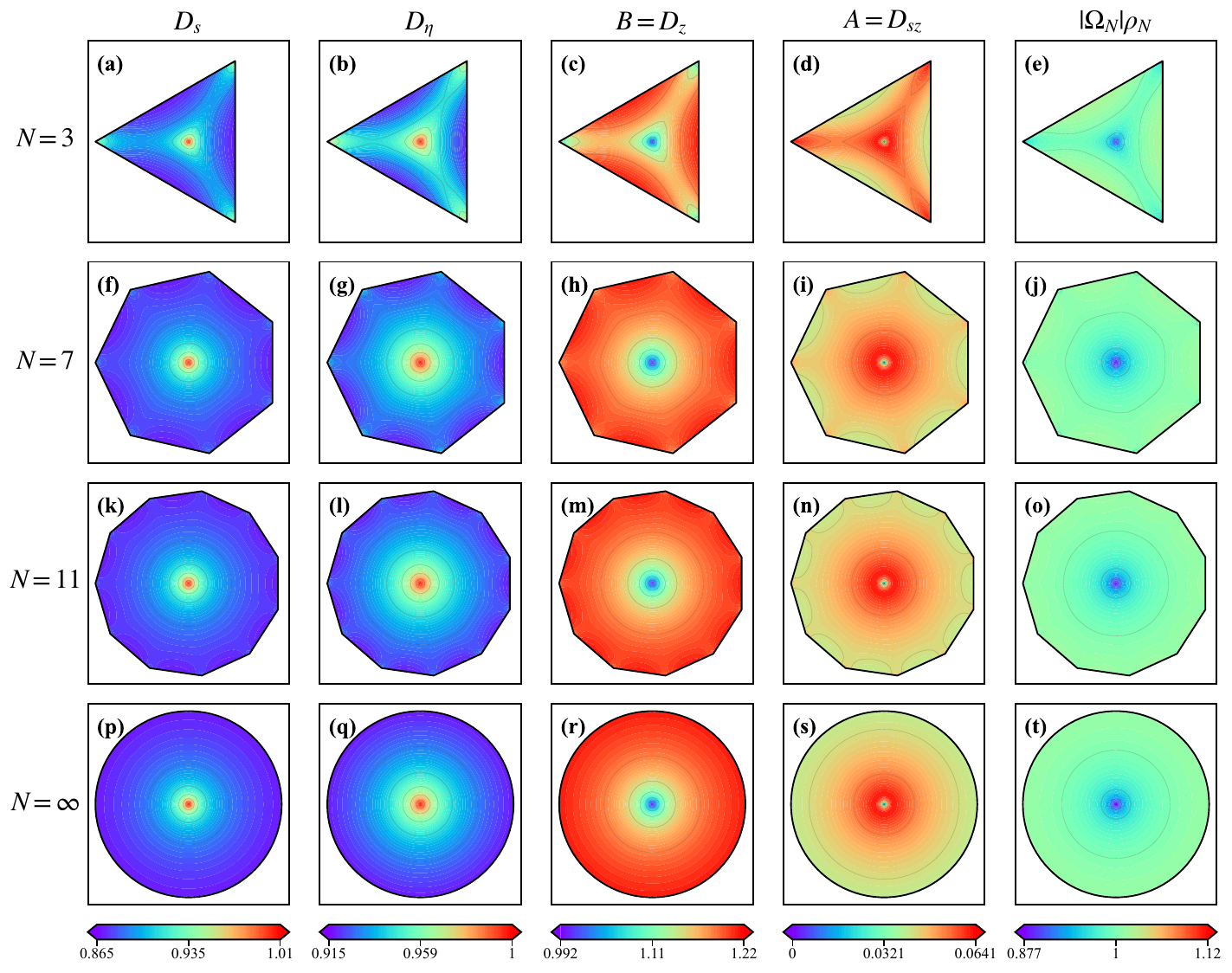}
  \caption{Cross-sectional transport fields for a larger aspect ratio and stronger alignment, \(p=100\) and \(\Per=50\). The rows correspond to \(N=3\), \(7\), \(11\) and \(\infty\). The columns are the same as in Figure~\ref{fig:fields-p10}.}
  \label{fig:fields-p100}
\end{figure}

\begin{figure}
  \centering
  \includegraphics[width=0.98\linewidth]{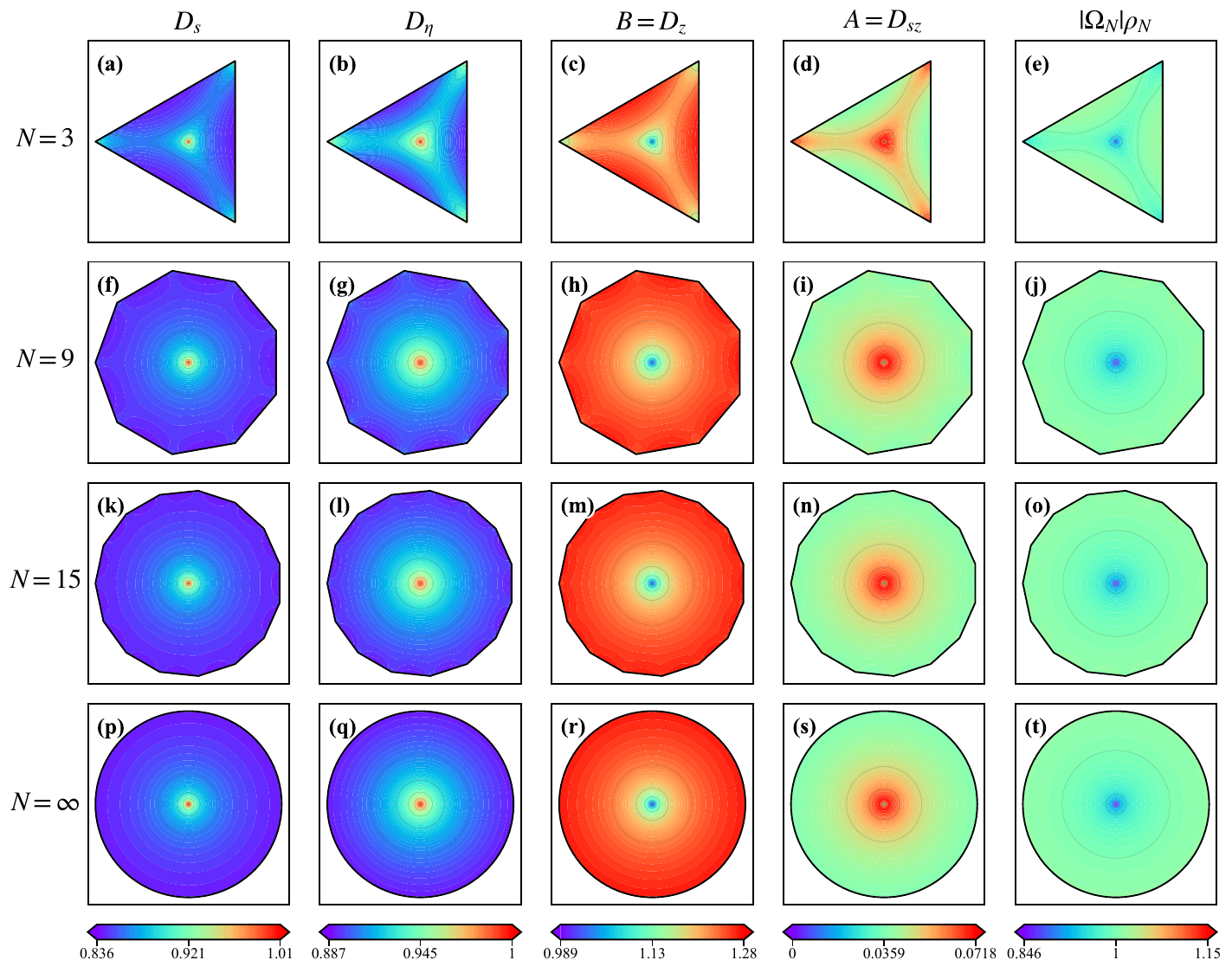}
  \caption{Cross-sectional transport fields near the slender-rod regime, \(p=1000\) and \(\Per=100\). The rows correspond to \(N=3\), \(9\), \(15\) and \(\infty\). The columns are \(D_s\), \(D_\eta\), \(B\), \(A\) and \(|\Omega_N|\rho_N\).}
  \label{fig:fields-p1000}
\end{figure}

Figures~\ref{fig:fields-p10}--\ref{fig:fields-p1000} show the spatial realization of the local curves in Figure~\ref{fig:local-closure} for the progression \((p,\Per)=(10,10)\), \((100,50)\) and \((1000,100)\). Each figure uses the same columns: \(D_s\), \(D_\eta\), \(B\), \(A\) and the scaled invariant density \(|\Omega_N|\rho_N\). The last scaling makes the spherical equilibrium equal to unity for every geometry, so departures from one show the redistribution caused by the conservative transverse operator.

These maps display a single sequence of effects. High-shear side-wall bands move the closure away from the isotropic branch: \(D_s\) and \(D_\eta\) decrease, \(B\) increases, and the mixed coefficient \(A\) appears where the orientational distribution is tilted but not yet fully streamwise. Because \(A(q)\) is non-monotone, its maxima do not necessarily coincide with the largest shear. The invariant density then responds to the transverse mobility fields rather than to the velocity field alone; regions of reduced transverse mobility acquire larger invariant weight, while weak-shear or fast-core regions may be depleted. These maps therefore show, before any Taylor reduction is performed, which cross-sectional regions carry the invariant mass at long times and which velocity contrasts will be available to the cell problem.

Figure~\ref{fig:fields-p10} gives a moderate-shear example in which the mechanism has begun but the coefficient contrast remains mild. The smallest values of \(D_s\) occur in the strongest side-wall shear, \(D_\eta\) varies more gently, and \(B\) is enhanced in the same aligned regions. The \(A\) field is displaced from the zero-shear interior and follows intermediate-shear regions, reflecting the non-monotone local curve in Figure~\ref{fig:local-closure}. The scaled density \(|\Omega_N|\rho_N\) already differs from one, showing that the transverse operator weights spatially varying mobility rather than area alone. For \(N=3\) the fields retain strong signatures of side-centred shear and weaker corner regions, whereas the circular row recovers an axisymmetric pattern.

Increasing the aspect ratio and the shear strength makes the tensor contrast and the finite-polygon geometry more visible (Figure~\ref{fig:fields-p100}). The transverse diffusion fields decrease substantially in the high-shear regions, while the axial component develops a stronger enhancement. The region of large \(A\) shifts toward intermediate-shear bands, because strongly aligned rods contribute little inclined moment once the distribution is nearly axial. The invariant density converts these coefficient gradients into enrichment and depletion regions: flat-wall shear bands, weak-shear corners and the fast interior are sampled with different equilibrium weights. The \(N=7\) and \(N=11\) rows are already close to the circular arrangement in the interior, while the triangular channel still keeps a visibly finite-polygon structure.

The near-slender case in Figure~\ref{fig:fields-p1000} shows the strongest redistribution among the three field maps. The transverse diffusivities approach their aligned-rod values in the largest-shear regions, while \(B\) is correspondingly increased. Regions of reduced transverse mobility acquire larger invariant weight in \(\rho_N\), so the long-time streamline sampling is shifted before any axial cell correction is applied. This redistribution is the origin of the mean-speed changes reported later. In particular, depletion of the weak-shear fast core and enrichment of slower high-shear regions provide the cross-sectional mechanism for the intermediate-\(\Per\) dip in \(\bar u_N\) discussed with the steady transport coefficients. The same transverse coefficients, together with the velocity deviation \(u_N-\bar u_N\), determine the cell problem and hence the Taylor coefficient. The high-\(N\) rows demonstrate polygon-to-pipe convergence at the level of the full tensor fields before cross-sectional averaging is performed.

\section{\texorpdfstring{Taylor--Aris reduction and effective transport coefficients}{Taylor-Aris reduction and effective transport coefficients}}
\label{sec:taylor-aris-reduction}

The conservative cross-sectional equation still contains the full polygonal coordinate \(\x\). The reduction follows the long-time Taylor--Aris and generalized-dispersion viewpoint, in which the axial field varies on a long scale while transverse relaxation remains \(O(1)\) \citep{taylor1953,aris1956,frankel1989foundations,ramirez2006,alexandre2021}. The orientation and transport fields are fixed by \((N,p,\Per)\), and the axial P{\'e}clet number is then taken in the asymptotic range \(\Pe\gg\Per\). The corresponding long-wave ordering is
\begin{equation}
  \Pe\,\partial_z C=O(C).
\label{eq:long-wave-ordering}
\end{equation}
The first cross-sectional correction may therefore be written as \(\Pe\chi_NC_z\), while the leading axial spreading enters the one-dimensional equation as \(\Pe^2\kappa_N\). The coefficient \(\kappa_N\) is a property of the fixed cross-sectional relaxation problem; once the regime \(\Pe\gg\Per\) is assumed, the axial P{\'e}clet number enters the leading Taylor diffusivity only through the prefactor \(\Pe^2\). Direct axial diffusion and the signed cross drift enter at lower order. The radial invariant measure of the circular tube is replaced here by the two-dimensional density \(\rho_N(\x)\).

\subsection{Cross-sectionally averaged concentration and mean speed}
\label{subsec:cross-section-average}

The one-dimensional concentration is the cross-sectional mass per unit axial length,
\begin{equation}
  C(z,t)
  =
  \int_{\Omega_N}c(\x,z,t)\,\dd A .
\label{eq:cross-section-C}
\end{equation}
Integrating \eqref{eq:polygon-conservation} over \(\Omega_N\) and using the wall condition \eqref{eq:polygon-wall-bc} gives the exact axial conservation law
\begin{equation}
  C_t+\partial_zF=0,
  \qquad
  F(z,t)=\int_{\Omega_N}J_z(\x,z,t)\,\dd A .
\label{eq:cross-section-flux}
\end{equation}
The Taylor--Aris reduction closes this conservation law by expressing \(F\) in terms of the slow axial field \(C\) \citep{aris1956,frankel1989foundations}.

At leading order the cross-section is in the invariant state found in Section~\ref{subsec:invariant-density},
\begin{equation}
  c_0(\x,z,t)=\rho_N(\x)C(z,t).
\label{eq:c0-leading-section4}
\end{equation}
Because \(\rho_N\) is normalized by \eqref{eq:rhoN-definition}, this form preserves the definition of \(C\). Substitution into the advective part of the axial flux gives
\begin{equation}
  F_0^{\rm adv}
  =
  \Pe
  \int_{\Omega_N}u_N(\x)\rho_N(\x)\,\dd A\, C .
\end{equation}
The leading sampling velocity is therefore
\begin{equation}
  \bar u_N
  =
  \int_{\Omega_N}u_N(\x)\rho_N(\x)\,\dd A .
\label{eq:ubarN-definition}
\end{equation}
Equation~\eqref{eq:ubarN-definition} defines the invariant-measure sampling speed. It is the streamline speed seen by the rod cloud after transverse relaxation under the alignment-modified mobility field. Redistribution of \(\rho_N\) shifts \(\bar u_N\) under the fixed Poiseuille velocity field. For spherical particles \(\rho_N=|\Omega_N|^{-1}\), so \(\bar u_N\) reduces to the area average of the centreline-normalized polygonal velocity.
At this order the one-dimensional equation is pure advection,
\begin{equation}
  C_t+\Pe\bar u_N C_z=0,
\label{eq:leading-1d-advection}
\end{equation}
and axial spreading enters only after the first cross-sectional correction is included.

\subsection{Cell problem}
\label{subsec:cell-problem}

Taylor dispersion is generated by the first departure from the invariant state. A slow axial gradient lets rods on faster-than-average and slower-than-average streamlines separate before transverse relaxation erases the contrast. We write the leading correction in the form
\begin{equation}
  c(\x,z,t)
  =
  \rho_N(\x)C(z,t)
  +
  \Pe\,\chi_N(\x)C_z(z,t)
  +\cdots ,
\label{eq:cell-expansion}
\end{equation}
with the mass constraint
\begin{equation}
  \int_{\Omega_N}\chi_N\,\dd A=0 .
\label{eq:chi-normalization}
\end{equation}
The factor \(\Pe\) records the strength of the axial velocity contrast in the nondimensional equation. Under the long-wave ordering \eqref{eq:long-wave-ordering}, \(\Pe\chi_NC_z\) remains the first correction to \(\rho_NC\), and \(\chi_N\) itself depends only on \((N,p,\Per)\).

Using the leading one-dimensional balance \(C_t+\Pe\bar u_NC_z=0\), the residual generated by \(c_0=\rho_NC\) in the advective equation is
\begin{equation}
  \Pe\,[u_N(\x)-\bar u_N]\rho_N(\x)C_z .
\end{equation}
This residual is balanced by transverse relaxation of the first correction, which gives the polygonal cell problem familiar from generalized Taylor-dispersion closures \citep{frankel1989foundations,ramirez2006,alexandre2021}:
\begin{subequations}
\label{eq:cell-problem-N}
\begin{align}
  \mathcal L_{0,N}\chi_N
  &=
  [u_N(\x)-\bar u_N]\rho_N(\x),
  \label{eq:cell-problem-interior}\\
  \bm n\cdot
  \left[
  \e_s\,\partial_s(D_s\chi_N)
  +
  \e_\eta\,\partial_\eta(D_\eta\chi_N)
  \right]
  &=0,
  \qquad
  \x\in\partial\Omega_N ,
  \label{eq:cell-problem-bc}\\
  \int_{\Omega_N}\chi_N\,\dd A&=0 .
  \label{eq:cell-problem-normalization}
\end{align}
\end{subequations}
The forcing in \eqref{eq:cell-problem-interior} is the invariantly weighted velocity fluctuation. The corrector \(\chi_N\) is the signed transverse memory generated by an imposed axial concentration gradient: faster-than-average and slower-than-average streamlines create opposite concentration distortions, and \(\mathcal L_{0,N}\) determines how strongly those distortions are relaxed. The solvability condition for \eqref{eq:cell-problem-N} is precisely \(\int_{\Omega_N}(u_N-\bar u_N)\rho_N\,\dd A=0\), so \(\bar u_N\) is fixed by the Fredholm compatibility condition for the cell problem.

\subsection{Taylor coefficient}
\label{subsec:taylor-coefficient}

The Taylor coefficient is the velocity-fluctuation forcing paired with the transverse relaxation response \citep{frankel1989foundations,ramirez2006,alexandre2021}. Substituting \eqref{eq:cell-expansion} into \(\Pe\int_{\Omega_N}u_Nc\,\dd A\) gives
\begin{equation}
  F^{\rm adv}
  =
  \Pe\bar u_NC
  +
  \Pe^2
  \left[
  \int_{\Omega_N}u_N\chi_N\,\dd A
  \right]C_z
  +\cdots .
\end{equation}
Because \(\chi_N\) has zero integral, the bracket can be written with \(u_N-\bar u_N\). We define
\begin{equation}
  \kappa_N
  =
  -
  \int_{\Omega_N}
  [u_N(\x)-\bar u_N]\chi_N(\x)\,\dd A .
\label{eq:kappaN-definition}
\end{equation}
The corresponding flux contribution is \(-\Pe^2\kappa_NC_z\), and therefore \(\Pe^2\kappa_NC_{zz}\) appears in the one-dimensional equation. Equation~\eqref{eq:kappaN-definition} is the operative polygonal definition: \(\chi_N\) is the zero-mass transverse response to the forcing \((u_N-\bar u_N)\rho_N\), and the pairing with \(u_N-\bar u_N\) gives the leading Taylor spreading coefficient.

When the transverse relaxation admits a reversible detailed-balance form, it is useful to introduce \(G_N=\chi_N/\rho_N\). In that case
\begin{equation}
  \mathcal L_{0,N}(\rho_NG_N)
  =
  \nabla_\perp\cdot
  \left(\bm K_N\rho_N\nabla_\perp G_N\right),
\end{equation}
and the Taylor coefficient has the energy representation
\begin{equation}
  \kappa_N
  =
  \int_{\Omega_N}
  \nabla_\perp G_N\cdot
  \bm K_N\rho_N\nabla_\perp G_N\,\dd A
  \ge0 .
\label{eq:kappaN-energy}
\end{equation}
For the generic polygonal product operator, \eqref{eq:kappaN-definition} is the coefficient used in the computations. The Dirichlet form \eqref{eq:kappaN-energy} records the reversible reduction and gives the circular-pipe expression below.

For comparisons across polygonal geometries the spherical reference must be computed in the same cross-section:
\begin{equation}
  \kappa_{s,N}
  =
  \kappa_N\big|_{p=1}.
\label{eq:kappa-sN}
\end{equation}
When \(p=1\), the coefficients are \(D_s=D_\eta=B=1\), \(A=0\), and \(\rho_N=|\Omega_N|^{-1}\). The velocity profile \(u_N\) and the domain \(\Omega_N\) still depend on \(N\), so the normalized enhancement reported below uses the same-geometry ratio \(\kappa_N/\kappa_{s,N}\). The circular value \(1/192\) is the reference only at \(N=\infty\).

\subsection{Direct axial diffusion and cross-diffusive drift}
\label{subsec:direct-cross-diffusion}

The tensor components \(B\) and \(A\) enter the reduced equation in different ways from the leading Taylor coefficient, as in orientable-particle and anisotropic-diffusion Taylor-dispersion formulations \citep{frankel1993,guan2024anisodiff}. The direct axial diffusion follows immediately from the term \(-Bc_z\) in \eqref{eq:polygon-axial-flux}. At leading cross-sectional equilibrium,
\begin{equation}
  \int_{\Omega_N}(-B c_z)\,\dd A
  =
  -
  \left[
  \int_{\Omega_N}B(\x)\rho_N(\x)\,\dd A
  \right]C_z ,
\end{equation}
so
\begin{equation}
  K_{{\rm dir},N}
  =
  \int_{\Omega_N}B(\x)\rho_N(\x)\,\dd A .
\label{eq:KdirN-definition}
\end{equation}
The coefficient \(K_{{\rm dir},N}\) is obtained by invariant-measure averaging of the local axial diffusivity. It records direct Brownian spreading along the duct and contributes \(O(1)\) to the unscaled axial diffusivity, subleading to \(\Pe^2\kappa_N\) in the high-\(\Pe\) Taylor scaling.

The cross coefficient \(A\) produces an advective correction through the conservative axial flux \(-\partial_s(Ac)\). Using \(c_0=\rho_NC\),
\begin{equation}
  \int_{\Omega_N}-\partial_s(A\rho_NC)\,\dd A
  =
  U_{A,N}C,
\end{equation}
where
\begin{equation}
  U_{A,N}
  =
  -
  \int_{\Omega_N}
  \e_s\cdot\nabla_\perp(A\rho_N)\,\dd A .
\label{eq:UAN-definition}
\end{equation}
Thus the off-diagonal component contributes to the mean migration speed at the next order:
\begin{equation}
  u_m
  =
  \bar u_N+\Pe^{-1}U_{A,N}+O(\Pe^{-2}),
\label{eq:umN-expansion}
\end{equation}
when the laboratory advection speed is written as \(\Pe u_m\). The quantity \(U_{A,N}\) is generated by spatial variation of the signed field \(A\rho_N\). A net migration correction requires a cross-sectional imbalance of this field along the shear direction, measured by the integral in \eqref{eq:UAN-definition}.
For spheres \(A=0\), and in the weak-orientational-shear limit \(A\to0\); hence \(U_{A,N}=0\) in both limits. Its sign for rods is tied to the convention \(\e_s=-\nabla_\perp u_N/|\nabla_\perp u_N|\) and to the signed definition of \(A\), which is why the shear-coordinate convention was fixed before the transport equation was introduced. The \(A\)-dependent transverse flux \(-\e_sAc_z\) belongs to the next cross-sectional correction in this ordering; the coefficient \(\kappa_N\) reported here is the leading advection--relaxation Taylor coefficient.

Collecting the retained terms gives the one-dimensional high-\(\Pe\) model in conservative form,
\begin{subequations}
\label{eq:effective1d-polygon}
\begin{align}
  C_t+\partial_zF_N[C]&=0,
  \label{eq:effective1d-conservation}\\
  F_N[C]
  &=
  V_N C-D_{{\rm eff},N}C_z,
  \label{eq:effective1d-flux}\\
  V_N&=\Pe\bar u_N+U_{A,N},
  \label{eq:effective1d-speed}\\
  D_{{\rm eff},N}
  &=\Pe^2\kappa_N+K_{{\rm dir},N}.
  \label{eq:effective1d-diffusivity}
\end{align}
\end{subequations}
Equivalently, for fixed \((N,p,\Per)\) so that the coefficients are independent of \(z\) and \(t\),
\begin{equation}
  C_t+V_NC_z=D_{{\rm eff},N}C_{zz}.
\label{eq:effective1d-advection-diffusion}
\end{equation}
Equations~\eqref{eq:effective1d-polygon}--\eqref{eq:effective1d-advection-diffusion} retain the leading \(O(\Pe^2)\) Taylor diffusivity together with the \(O(1)\) direct-diffusion and cross-drift corrections. The three cross-sectional objects \(\rho_N\), \(\chi_N\) and \(A\rho_N\) determine, respectively, the leading velocity sampling, the leading Taylor dispersion, and the conservative drift correction.
The steady coefficients retained in the reduced model are
\begin{equation}
  \bar u_N,
  \qquad
  \frac{\kappa_N}{\kappa_{s,N}},
  \qquad
  U_{A,N}.
\label{eq:steady-output-summary}
\end{equation}
The steady-result discussion below uses \(\bar u_N\) and \(\kappa_N/\kappa_{s,N}\) as the leading high-\(\Pe\) diagnostics, and reports \(U_{A,N}\) as a lower-order signed drift coefficient.

\subsection{Circular-pipe reference branch}
\label{subsec:circular-reference-branch}

The circular limit provides both a check against the classical circular-tube result \citep{taylor1953,aris1956} and against the tensorial Brownian-rod tube reduction \citep{feng2026tube}. It is also the reference branch used for the finite-\(N\) convergence. For \(N=\infty\),
\begin{equation}
  \Omega_\infty=\{\x:|\x|<1\},
  \qquad
  u_\infty(r)=1-r^2,
  \qquad
  q_\infty(r;\Per)=\Per r .
\end{equation}
With the sign convention \(\e_s=\e_r\), define
\begin{equation}
  D(r)=D_s(\Per r;p),
  \qquad
  A(r)=A(\Per r;p),
  \qquad
  B(r)=B(\Per r;p).
\label{eq:circular-DAB}
\end{equation}
Axisymmetric concentrations have no \(\eta\)-derivative, so \(D_\eta\) drops out of the radial reduction. The invariant density is
\begin{equation}
  \rho_\infty(r)
  =
  \frac{1}{2\pi I_0D(r)},
  \qquad
  I_0=\int_0^1rD^{-1}(r)\,\dd r .
\label{eq:rho-pipe-reference}
\end{equation}
Consequently
\begin{equation}
  \bar u_\infty
  =
  \frac{
  \displaystyle\int_0^1(1-r^2)rD^{-1}(r)\,\dd r
  }{
  \displaystyle\int_0^1rD^{-1}(r)\,\dd r
  } .
\label{eq:ubar-pipe-reference}
\end{equation}

Writing the circular corrector as
\begin{equation}
  \chi_\infty(r)=\frac{G(r)}{2\pi I_0D(r)}
\end{equation}
reduces \eqref{eq:cell-problem-N} to
\begin{equation}
  \frac1r\frac{\dd}{\dd r}
  \left(rG'(r)\right)
  =
  \frac{u_\infty(r)-\bar u_\infty}{D(r)},
  \qquad
  rG'(r)\to0\ (r\to0),\quad G'(1)=0 .
\label{eq:circular-G-problem}
\end{equation}
The additive constant in \(G\) may be fixed by \(\int_0^1rG(r)D^{-1}(r)\,\dd r=0\). The Taylor coefficient then has the positive form
\begin{equation}
  \kappa_\infty
  =
  \frac{
  \displaystyle\int_0^1 r[G'(r)]^2\,\dd r
  }{
  \displaystyle\int_0^1 rD^{-1}(r)\,\dd r
  } .
\label{eq:kappa-pipe-reference}
\end{equation}
The remaining lower-order coefficients reduce to
\begin{equation}
  K_{{\rm dir},\infty}
  =
  \frac{
  \displaystyle\int_0^1 rB(r)D^{-1}(r)\,\dd r
  }{
  \displaystyle\int_0^1 rD^{-1}(r)\,\dd r
  },
\label{eq:Kdir-pipe-reference}
\end{equation}
and, with the same radial--axial sign convention for \(A\) implied by \(\e_s=\e_r\),
\begin{equation}
  U_{A,\infty}
  =
  \frac{
  \displaystyle\int_0^1 A(r)D^{-1}(r)\,\dd r
  -
  A(1)D^{-1}(1)
  }{
  \displaystyle\int_0^1 rD^{-1}(r)\,\dd r
  } .
\label{eq:UA-pipe-reference}
\end{equation}
This follows directly from \(U_{A,\infty}=-2\pi\int_0^1 r\,\frac{\dd}{\dd r}(A\rho_\infty)\,\dd r\), and it vanishes when \(A\) and \(D\) are constant.
For spherical particles \(D\equiv1\), \(A=0\), \(B=1\), and the classical centreline-normalized circular-tube values are recovered:
\begin{equation}
  \bar u_\infty=\frac12,
  \qquad
  U_{A,\infty}=0,
  \qquad
  K_{{\rm dir},\infty}=1,
  \qquad
  \kappa_{s,\infty}=\frac1{192}.
\label{eq:circular-spherical-reference}
\end{equation}
The finite-polygon results should approach this branch as \(N\to\infty\), while comparisons at fixed finite \(N\) use the same-geometry spherical coefficient \(\kappa_{s,N}\) from \eqref{eq:kappa-sN}.

\section{Steady Taylor--Aris coefficients in polygonal ducts}
\label{sec:steady-effective-transport-results}

The steady coefficients reveal two distinct consequences of Jeffery--Brownian alignment, consistent with earlier orientable-particle and Brownian-rod Taylor-dispersion studies \citep{frankel1993,kumar2021,khair2022}. First, alignment modifies the invariant cross-sectional sampling and therefore produces a small, non-monotone shift in the leading mean speed \(\bar u_N\). Second, and more strongly, it reduces transverse relaxation in the Taylor cell problem, producing a monotone enhancement of the leading Taylor coefficient. The same-geometry normalization \(R_{\kappa,N}=\kappa_N/\kappa_{s,N}\) is used to isolate this rod-induced enhancement from the passive dependence of the Poiseuille cell problem on polygonal shape.

We denote by \(\kappa_{m,N}\) the fully aligned same-geometry reference coefficient. The normalized enhancement used below is
\begin{equation}
  E_N(\Per,p)=
  \frac{\kappa_N/\kappa_{s,N}-1}
       {\kappa_{m,N}/\kappa_{s,N}-1},
\label{eq:EN-definition}
\end{equation}
for \(p>1\), with the spherical branch shown as the baseline \(E_N=0\).

Figures~\ref{fig:mean-speed-polygons}--\ref{fig:enhancement-collapse-polygons} quantify the dependence on \(\Per\), \(p\) and finite polygon geometry. Figure~\ref{fig:polygon-pipe-convergence} then checks convergence of the finite-polygon coefficients to the circular-pipe branch. The lower-order coefficients \(U_{A,N}\) and \(K_{{\rm dir},N}\) are retained in the reduced equation \eqref{eq:effective1d-polygon}; \(U_{A,N}\) is summarized below as a signed drift diagnostic, while the main steady-result discussion focuses on the leading high-\(\Pe\) sampling and Taylor-dispersion coefficients. The direct diffusivity \(K_{{\rm dir},N}\) is not tabulated because it is an invariant-measure average of \(B(\x)\) and remains an \(O(1)\) additive contribution to \(D_{{\rm eff},N}\), whereas the leading high-\(\Pe\) variation of axial spreading is controlled by \(\Pe^2\kappa_N\).

\subsection{Mean transport speed}
\label{subsec:mean-transport-speed}

\begin{figure}
  \centering
  \includegraphics[width=0.98\linewidth]{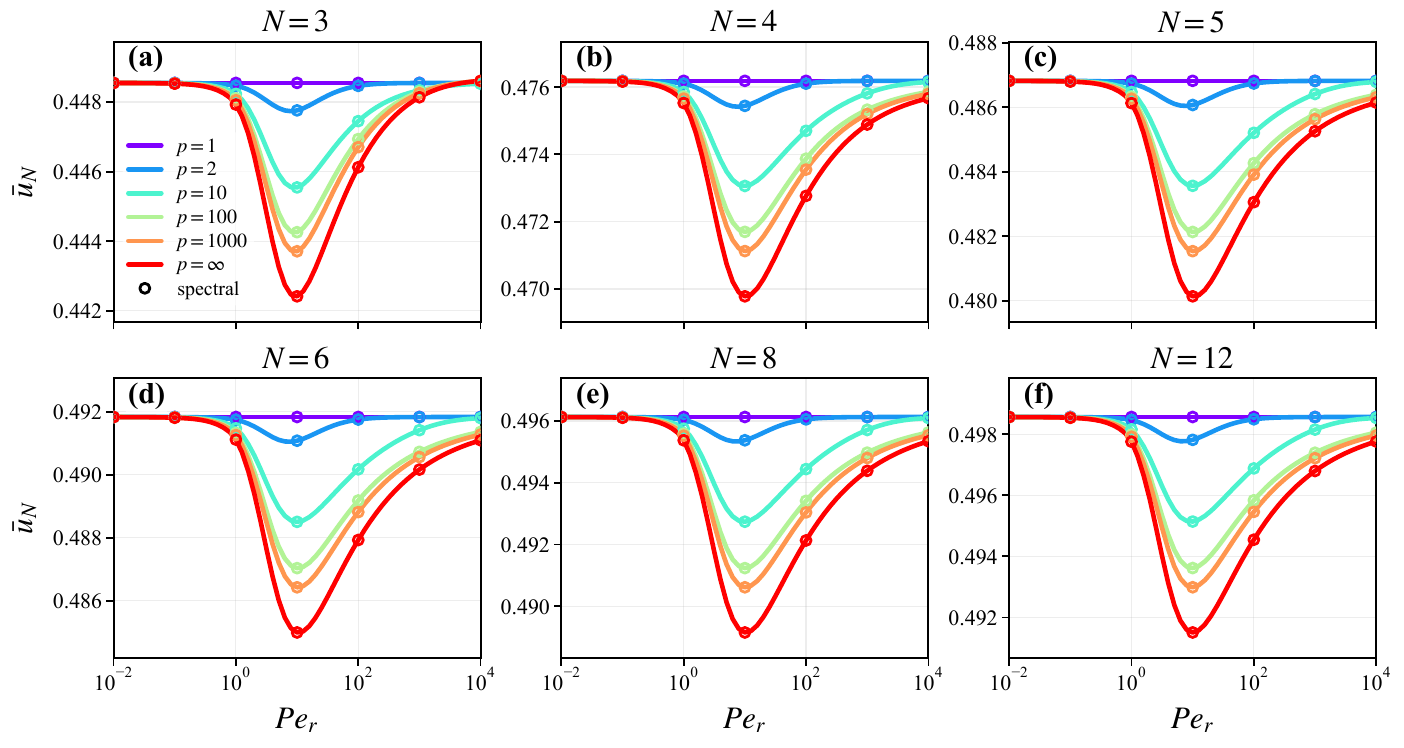}
  \caption{Leading mean transport speed \(\bar u_N\) as a function of the rotational P{\'e}clet number \(\Per\). Each panel corresponds to one regular polygon, \(N=3,4,5,6,8,12\). Curves show aspect ratios \(p=1,2,10,100,1000,\infty\). Open markers show independent evaluations at selected \(\Per\) values using the spectral discretization described in Section~\ref{sec:transient-spectral-validation}.}
  \label{fig:mean-speed-polygons}
\end{figure}

The spherical branch gives the geometric baseline. For \(p=1\), the local closure is isotropic and the invariant density is uniform, so \(\bar u_N\) is simply the area average of the centreline-normalized Poiseuille profile. The curves are therefore horizontal in \(\Per\). Their values increase from \(0.44855\) for the triangular channel to \(0.49857\) for \(N=12\), approaching the circular value \(1/2\). This variation is not a rod effect; it is the passive consequence of using a fixed centreline velocity scale in different cross-sections \citep{shah1975,shah1978,tamayol2010}.

For \(p>1\), the mean speed develops a shallow non-monotone dependence on \(\Per\). The mechanism is the invariant-density redistribution already visible in Figures~\ref{fig:fields-p10}--\ref{fig:fields-p1000}. At weak orientational shear, the closure is nearly isotropic and the density is nearly uniform. At intermediate \(\Per\), alignment first suppresses transverse mobility in high-shear regions near the walls, where the Poiseuille speed is lower than in the core. The product-form transverse operator then favours reduced-mobility regions in its invariant state: in the circular limit this weighting reduces exactly to \(D_s^{-1}\), while in polygons the same mechanism gives a two-dimensional analogue controlled by the spatially varying \(D_s\) and \(D_\eta\) fields and by the non-global shear frame. The rod cloud therefore samples slower streamlines more often, producing the dip in \(\bar u_N\).

The dip occurs at \(\Per=O(10)\) for the moderate and slender rods. In the \(N=12\) panel, the minimum falls from the spherical value \(0.49857\) to \(0.49514\) for \(p=10\), \(0.49299\) for \(p=1000\), and \(0.49151\) in the infinitely slender limit. Thus even the strongest mean-speed change is only about \(1.4\%\) of the same-geometry spherical value. The small magnitude of the dip indicates that \(\bar u_N\) is primarily a diagnostic of invariant-density redistribution, rather than the dominant steady high-\(\Pe\) signature in the coefficients plotted here.

At larger \(\Per\), the aligned region spreads over most of the finite-shear area. The transverse-mobility contrast that originally produced the invariant-density shift is then reduced, except near weak-shear neighbourhoods, and the sampling velocity moves back toward the spherical baseline. For \(N=12\), the \(p=1000\) curve has returned to \(0.49796\) by \(\Per=10^4\), very close to the spherical baseline. This recovery distinguishes \(\bar u_N\) from the Taylor coefficient: the mean speed records where the invariant measure places mass, whereas \(\kappa_N\) remains sensitive to the reduced transverse relaxation in the cell problem.

\begin{table}
  \centering
  \caption{Representative steady coefficients from Figures~\ref{fig:mean-speed-polygons}--\ref{fig:enhancement-collapse-polygons} for \(N=12\). The second and third columns give the minimum of \(\bar u_N\) over the plotted \(\Per\) range and the corresponding \(\Per\). The fourth, fifth and sixth columns report \(\bar u_N\), \(R_{\kappa,N}\) and \(E_N\) at the high-shear endpoint \(\Per=10^4\).}
  \label{tab:steady-summary-N12}
  \begin{tabular}{cccccc}
    \toprule
    \(p\) & \(\bar u_N^{\min}\) & \(\Per\) at min & \(\bar u_N\) & \(R_{\kappa,N}\) & \(E_N\) \\
    \midrule
    \(1\) & 0.49857 & -- & 0.49857 & 1.000 & -- \\
    \(2\) & 0.49777 & 6.31 & 0.49857 & 1.017 & 0.357 \\
    \(10\) & 0.49514 & 10 & 0.49853 & 1.119 & 0.810 \\
    \(100\) & 0.49362 & 10 & 0.49805 & 1.200 & 0.914 \\
    \(1000\) & 0.49299 & 10 & 0.49796 & 1.229 & 0.914 \\
    \(\infty\) & 0.49151 & 10 & 0.49776 & 1.303 & 0.909 \\
    \bottomrule
  \end{tabular}
\end{table}

The signed drift coefficient \(U_{A,N}\) is much smaller in the laboratory migration speed because it enters \(V_N=\Pe\bar u_N+U_{A,N}\). Table~\ref{tab:cross-drift-summary-N12} reports the same \(N=12\) data used in Figures~\ref{fig:mean-speed-polygons}--\ref{fig:enhancement-collapse-polygons}. The coefficient vanishes for spheres, changes sign for rods as the non-monotone local \(A(q)\) field moves across the section, and remains \(O(10^{-2})\) over the plotted range. As a conservative separation diagnostic, the fourth column evaluates the largest relative contribution obtained by taking the minimal pointwise separation \(\Pe=10\Per\); any larger axial P{\'e}clet number reduces this ratio in proportion to \(1/\Pe\).

\begin{table}
  \centering
  \caption{Representative lower-order cross-diffusive drift for \(N=12\). The extrema are taken over the plotted \(\Per\) range. The fourth column reports the largest value of \(|U_{A,N}|/(\Pe\bar u_N)\) over the same range under the minimal separation choice \(\Pe=10\Per\). The final column gives the drift coefficient at the high-shear endpoint \(\Per=10^4\). The sign follows the convention \(\e_s=-\nabla_\perp u_N/|\nabla_\perp u_N|\).}
  \label{tab:cross-drift-summary-N12}
  \begin{tabular}{ccccc}
    \toprule
    \(p\) & \(U_{A,N}^{\min}\) \((\Per)\) & \(U_{A,N}^{\max}\) \((\Per)\) & \(\max |U_{A,N}|/(10\Per\bar u_N)\) & \(U_{A,N}\) at \(\Per=10^4\) \\
    \midrule
    \(1\) & \(0\) & \(0\) & \(0\) & \(0\) \\
    \(2\) & \(-0.00602\) \((2.00)\) & \(0.00456\) \((31.6)\) & \(1.07\times10^{-3}\) & \(5.12\times10^{-5}\) \\
    \(10\) & \(-0.0287\) \((2.51)\) & \(0.0150\) \((126)\) & \(4.82\times10^{-3}\) & \(0.00262\) \\
    \(100\) & \(-0.0416\) \((2.51)\) & \(0.0167\) \((126)\) & \(6.89\times10^{-3}\) & \(0.00578\) \\
    \(1000\) & \(-0.0468\) \((2.51)\) & \(0.0184\) \((126)\) & \(7.71\times10^{-3}\) & \(0.00628\) \\
    \(\infty\) & \(-0.0589\) \((2.51)\) & \(0.0222\) \((158)\) & \(9.60\times10^{-3}\) & \(0.00776\) \\
    \bottomrule
  \end{tabular}
\end{table}

\subsection{Taylor dispersion enhancement}
\label{subsec:taylor-dispersion-enhancement}

\begin{figure}
  \centering
  \includegraphics[width=0.98\linewidth]{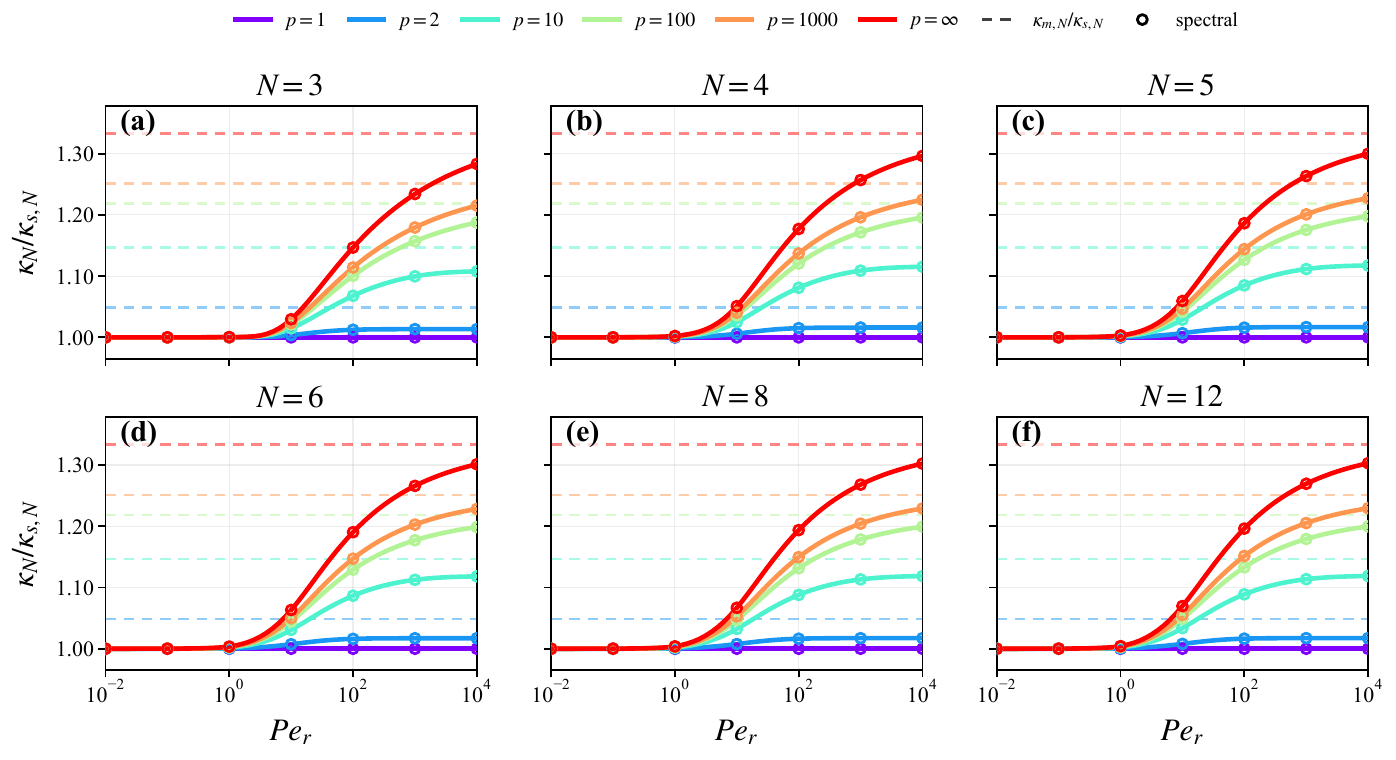}
  \caption{Taylor-dispersion enhancement \(R_{\kappa,N}=\kappa_N/\kappa_{s,N}\) as a function of \(\Per\). The panel layout and aspect-ratio curves are the same as in Figure~\ref{fig:mean-speed-polygons}. Dashed lines indicate the fully aligned reference \(1/d_\perp(p)\). Open markers show independent evaluations at selected \(\Per\) values using the spectral discretization described in Section~\ref{sec:transient-spectral-validation}.}
  \label{fig:taylor-enhancement-polygons}
\end{figure}

The Taylor coefficient responds much more strongly than the mean speed. In the weak-shear limit the local closure approaches the isotropic branch, so all curves start near \(R_{\kappa,N}=1\). As \(\Per\) increases, the high-shear regions become streamwise aligned. The transverse components \(D_s\) and \(D_\eta\) then fall below the isotropic value, cross-sectional exchange slows, and the cell correction can maintain a larger velocity deviation before transverse relaxation removes it. This is the same alignment-controlled mechanism identified in simpler rod-dispersion settings \citep{frankel1993,kumar2021,khair2022}, but here it acts through the full two-dimensional density and tensor field rather than through a radial coefficient alone.

The same-geometry normalization removes most of the passive dependence on \(N\). For \(p=1000\) and \(\Per=10^4\), \(R_{\kappa,N}\) increases only from \(1.215\) at \(N=3\) to \(1.229\) at \(N=12\), even though the unnormalized spherical Taylor coefficient changes substantially with polygon shape. After this normalization, most of the remaining variation is controlled by the local orientation closure. Polygonal geometry still enters through the distribution of \(q_N(\x)\), the invariant density and the cell problem, but its residual effect on the normalized enhancement is comparatively weak for the cases shown.

The dashed lines in Figure~\ref{fig:taylor-enhancement-polygons} show the fully aligned transverse-mixing limits \(1/d_\perp(p)\), with \(d_\perp(p)\) set by Perrin's transverse diffusivity for a prolate spheroid \citep{perrin1936}. In this ideal limit the transverse relaxation operator is approximately scaled by the uniform transverse diffusivity \(d_\perp(p)\). The cell response therefore scales as \(d_\perp^{-1}\), giving \(\kappa_{m,N}/\kappa_{s,N}=1/d_\perp(p)\) when the velocity field and geometry are held fixed. These limits are \(1.048\) for \(p=2\), \(1.147\) for \(p=10\), \(1.218\) for \(p=100\), \(1.251\) for \(p=1000\), and \(4/3\) in the infinitely slender limit. The end points at \(\Per=10^4\) are still below the corresponding limits. For \(N=12\), Table~\ref{tab:steady-summary-N12} gives \(R_{\kappa,N}=1.229\) for \(p=1000\) and \(1.303\) for \(p\to\infty\). The remaining gap reflects finite-\(q\) orientational diffusion and the weak-shear parts of the section, where the distribution cannot be fully axial.

The dependence on \(p\) follows from the available diffusivity contrast \(d_\parallel-d_\perp\). Short rods have a small anisotropy in translational diffusion, so the available enhancement interval is small and the \(p=2\) curve stays close to unity. For \(p\ge10\), the longitudinal and transverse diffusivities are sufficiently separated that shear alignment gives an appreciable Taylor response. Within the leading high-\(\Pe\) coefficients plotted here, the increase of \(R_{\kappa,N}\) is the dominant steady signature of rod alignment, whereas the dip in \(\bar u_N\) mainly records the accompanying shift in invariant sampling.

\subsection{Normalized enhancement collapse}
\label{subsec:normalized-enhancement-collapse}

\begin{figure}
  \centering
  \includegraphics[width=0.78\linewidth]{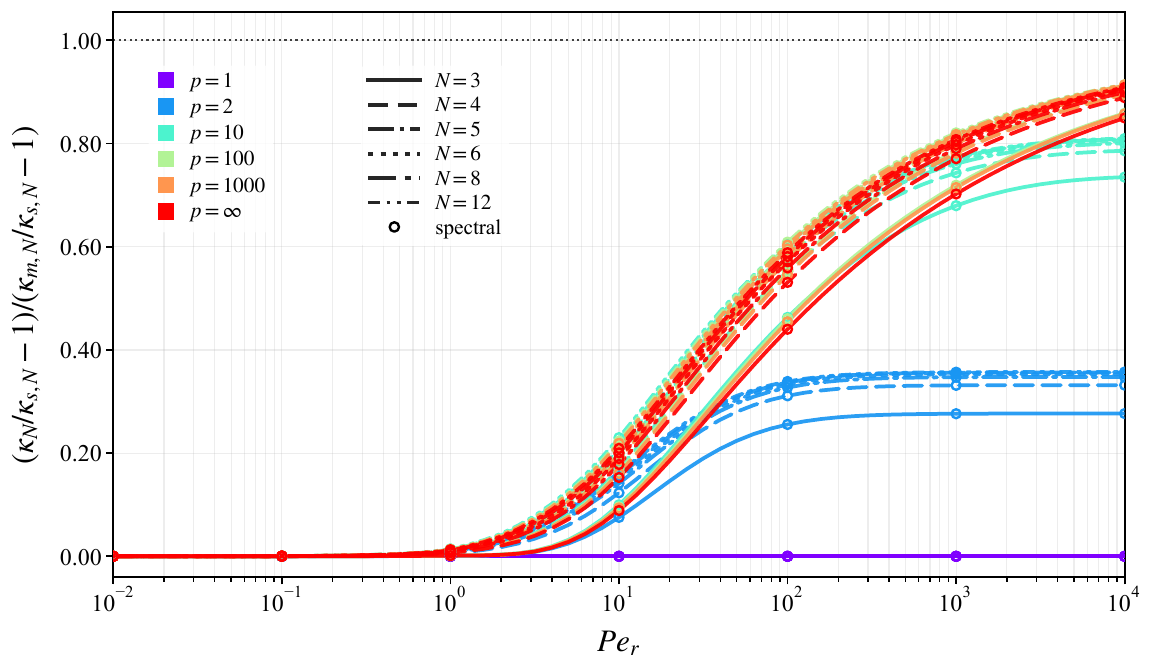}
  \caption{Normalized Taylor-dispersion enhancement
  \(E_N=(\kappa_N/\kappa_{s,N}-1)/(\kappa_{m,N}/\kappa_{s,N}-1)\). Curves collect the polygonal data from Figure~\ref{fig:taylor-enhancement-polygons}. The normalization removes the aspect-ratio-dependent fully aligned scale; \(E_N=1\) is the fully aligned reference.}
  \label{fig:enhancement-collapse-polygons}
\end{figure}

The aspect-ratio dependence of the enhancement separates naturally into an amplitude and an approach to the aligned state. In the present calculation the reference in \eqref{eq:EN-definition} satisfies \(\kappa_{m,N}/\kappa_{s,N}=1/d_\perp(p)\), which is why the denominator is the fully aligned enhancement interval indicated by the dashed lines in Figure~\ref{fig:taylor-enhancement-polygons}.

Figure~\ref{fig:enhancement-collapse-polygons} shows a strong collapse for moderate and slender rods once the enhancement is normalized by the aligned-state amplitude. The normalization separates two effects: \(d_\perp(p)\) sets the available enhancement amplitude, whereas \(\Per\) controls how much of the cross-section has entered the aligned branch of the local closure. The collapse is weaker for \(p=2\), where the Jeffery bias and the translational anisotropy are both small.

The residual deviations identify the limits of this aligned-amplitude scaling. For \(p=2\), the weak Jeffery bias delays the approach to the aligned branch. For \(p\ge100\), the remaining distance from \(E_N=1\) is mainly a finite-\(\Per\) correction of the local orientation closure, not a finite-\(N\) error. In the \(N=12\) data summarized in Table~\ref{tab:steady-summary-N12}, \(E_N(10^4)\) is about \(0.91\) for \(p=100\), \(p=1000\) and \(p\to\infty\). Polygonal geometry produces additional spread because different cross-sections distribute the same maximum-normalized shear over the area in different ways, but this spread is secondary after the aligned-scale normalization.

The normalization separates the velocity-sampling effect from the transverse-relaxation effect. The mean speed responds to where the invariant density places mass in the velocity profile and can recover once the density becomes nearly uniform again. The Taylor coefficient responds to the transverse mixing time in the cell problem. Even when \(\bar u_N\) has nearly returned to its spherical value at high \(\Per\), \(E_N\) can remain close to one because the transverse diffusivity is still close to the aligned value that controls the cell response.

\subsection{Polygon-to-pipe convergence}
\label{subsec:polygon-pipe-convergence}

\begin{figure}
  \centering
  \includegraphics[width=0.98\linewidth]{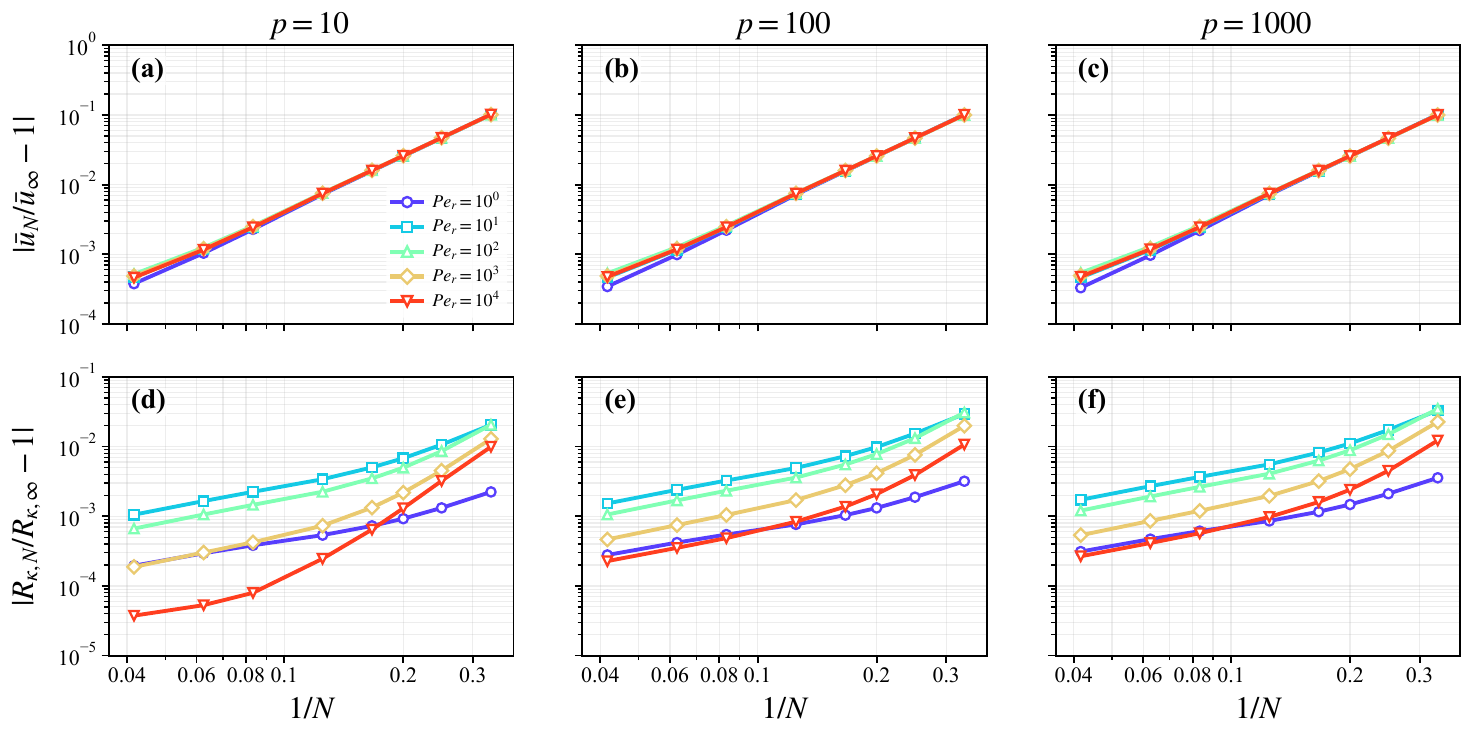}
  \caption{Finite-\(N\) convergence of steady transport coefficients to the circular-pipe branch. The upper row shows \(|\bar u_N/\bar u_\infty-1|\). The lower row shows \(|R_{\kappa,N}/R_{\kappa,\infty}-1|\), where \(R_{\kappa,N}=\kappa_N/\kappa_{s,N}\). Columns correspond to \(p=10,100,1000\), and curves correspond to \(\Per=1,10,100,1000,10000\).}
  \label{fig:polygon-pipe-convergence}
\end{figure}

The circular formulas in Section~\ref{subsec:circular-reference-branch} provide the limiting branch for the polygonal computations. Figure~\ref{fig:polygon-pipe-convergence} uses additional calculations at \(N=16\) and \(N=24\), together with the circular branch, to measure finite-\(N\) errors in the steady invariant-density and cell-problem coefficients. No asymptotic convergence rate is inferred here; the figure quantifies the finite-\(N\) error over the computed range.

The upper row shows that the leading mean speed converges rapidly to the circular value. The triangular channel is the outlier: across the cases in Figure~\ref{fig:polygon-pipe-convergence}, its largest relative deviation in \(\bar u_N\) is about \(10\%\). This large number mostly reflects passive geometric differences in the centreline-normalized Poiseuille profile. By \(N=6\) the maximum relative deviation is about \(1.6\%\), by \(N=12\) it is about \(2.6\times10^{-3}\), and by \(N=24\) it is below \(5.5\times10^{-4}\).

The lower row gives the corresponding convergence of the rod-induced Taylor enhancement ratio. Because \(R_{\kappa,N}\) has already been normalized by the same-geometry spherical coefficient, the passive geometric part is largely removed. The triangular channel still has the largest deviation from the pipe branch, but the maximum is only about \(3.5\%\) over the cases plotted. The largest deviations for \(N=12\) and \(N=24\) are about \(3.7\times10^{-3}\) and \(1.7\times10^{-3}\), respectively. Thus the Taylor-enhancement ratio reaches the pipe branch smoothly as the polygon approaches the disk.

The \(\Per\)-dependence of the finite-\(N\) correction follows the same physical balance as the preceding panels. At weak orientational shear, the closure is close to isotropic and finite-\(N\) differences are mostly passive. At intermediate \(\Per\), density redistribution and Taylor enhancement are both active, so the details of the polygonal shear field matter most for \(R_{\kappa,N}\). At very large \(\Per\), much of the section is close to the aligned transverse diffusivity, and the normalized enhancement becomes less sensitive to the exact polygonal distribution of shear. The convergence test therefore supports two uses of the theory: small \(N\) can be treated as genuinely polygonal geometries, while \(N\gtrsim12\) already gives a close approximation to the circular-pipe branch for the steady coefficients. This statement concerns only \(\bar u_N\) and \(R_{\kappa,N}\); transient modal spectra and injection-memory effects need not converge at the same rate.

The steady coefficients therefore separate the role of alignment into two mechanisms. The invariant density produces a small, non-monotone shift in the sampled mean speed by redistributing mass among streamlines. The Taylor cell problem produces a larger, monotone increase in axial dispersion because streamwise alignment reduces transverse relaxation. After normalization by the same-geometry spherical coefficient and by the aligned transverse-diffusivity scale, the remaining finite-polygon correction is controlled mainly by how \(q_N(\x)\) is distributed over the cross-section and vanishes as \(N\to\infty\), recovering the circular-tube tensorial branch \citep{feng2026tube}.

\section{Transient relaxation to the Taylor--Aris regime}
\label{sec:transient-spectral-validation}

Section~\ref{sec:taylor-aris-reduction} eliminates the cross-section after transverse equilibration and yields the cell-problem coefficient \(\kappa_N\). A finite injection begins from a transverse profile that may differ strongly from the invariant density. The finite-time question is how this initial transverse memory decays and how the axial variance selects the Taylor--Aris coefficient. We address this question by evolving the uneliminated high-\(\Pe\) advection--relaxation equation in a transverse eigenbasis. The zero right mode is the invariant density \(\rho_N\); the non-zero modes are injection-dependent transverse structures. Their eigenvalues set relaxation times, and their coupling to the Poiseuille velocity determines their contribution to transient axial spreading \citep{vedel2012,vedel2014,jiang2021,feng2026tube}.

In this section \(\kappa_{{\rm TA},N}\) denotes the same leading coefficient written as \(\kappa_N\) in Sections~\ref{sec:taylor-aris-reduction} and \ref{sec:steady-effective-transport-results}. The extra subscript distinguishes the long-time Taylor--Aris reference value from finite-time running estimates. The calculation below tests convergence to this limit within the leading high-\(\Pe\) advection--relaxation operator. Appendix~\ref{app:full-fourier-fem-validation} gives an independent full transverse-space Fourier--FEM small-wavenumber validation of the reduced spectral coefficients.

The open markers in Figures~\ref{fig:mean-speed-polygons} and \ref{fig:taylor-enhancement-polygons} were generated from the steady zero-mode and cell-problem components of this spectral discretization. Their overlap with the curves in Section~\ref{sec:steady-effective-transport-results} checks that the spectral implementation recovers the steady sampling speed and Taylor coefficient before it is used for the finite-time calculations below.

Unless otherwise stated, the transient calculations use \(p=100\), \(\Pe=10^4\), \(\Per=10\) and \(N=3,4,5,7\). Time is reported in the scaled form
\begin{equation}
  t^\ast=\lambda_1t,
\end{equation}
where \(\lambda_1\) is the first non-zero transverse relaxation eigenvalue for the corresponding polygon. This scaling places the modal relaxation histories for different \(N\) on a common transverse-mixing time scale.
Equal values of \(t^\ast\) therefore represent equal fractions of the slowest transverse relaxation time, while the corresponding dimensional time is \(t=t^\ast/\lambda_1\) and varies with \(N\).

\subsection{Spectral transient setup}
\label{subsec:spectral-transient-setup}

The leading transient model keeps axial advection by the Poiseuille profile and transverse relaxation by \(\mathcal L_{0,N}\):
\begin{equation}
  c_t+\Pe u_N(\x)c_z=\mathcal L_{0,N}c,
  \qquad \x\in\Omega_N .
\label{eq:polygon-leading-transient-model}
\end{equation}
It uses the conservative no-flux boundary condition associated with \(\mathcal L_{0,N}\), as in \eqref{eq:rhoN-definition}. This is the same transverse mixing law that determines \(\rho_N\) and the cell problem \citep{aris1956,frankel1989foundations,ramirez2006,alexandre2021}. The model deliberately retains the leading high-\(\Pe\) advection--relaxation balance. The omitted direct axial diffusion contributes \(K_{{\rm dir},N}\) to \(D_{{\rm eff},N}\), which appears as \(K_{{\rm dir},N}/\Pe^2\) in the normalized running coefficient. The conservative \(A\)-dependent term contributes the \(O(1)\) drift correction \(U_{A,N}\) to the laboratory migration speed, compared with the leading \(O(\Pe)\) advection. In the scaling of Figure~\ref{fig:transient-running-kappa},
\[
  \frac{K_{{\rm dir},N}}{\Pe^2\kappa_{{\rm TA},N}}=O(\Pe^{-2}),
  \qquad
  \frac{U_{A,N}}{\Pe\bar u_N}=O(\Pe^{-1}).
\]
At \(\Pe=10^4\), these lower-order terms leave the leading convergence of the running coefficient to \(\kappa_{{\rm TA},N}\) unchanged.

The circular-pipe reduction admits a self-adjoint Sturm--Liouville form. In a polygon, the directed tensor fields and conservative coefficient placement give a generally non-self-adjoint transverse operator. We therefore use both right and left modes. Let \(\bm A_0\) denote the discrete positive relaxation matrix associated with \(-\mathcal L_{0,N}\), and let \(\bm M\) be the mass matrix. The retained modes satisfy
\begin{equation}
  \bm A_0\bm R_m=\lambda_m\bm M\bm R_m,
  \qquad
  \bm A_0^T\bm L_m=\lambda_m\bm M^T\bm L_m,
\label{eq:polygon-spectral-eigenproblem}
\end{equation}
with biorthogonal normalization
\begin{equation}
  \bm L_m^\ast\bm M\bm R_n=\delta_{mn}.
\label{eq:polygon-biorthogonality}
\end{equation}
Here and below, the asterisk denotes conjugate transpose.
For the computed cases the retained relaxation eigenvalues are real to numerical tolerance. If a parameter set produces complex conjugate pairs, the relaxation rates are interpreted through the positive real parts of \(\lambda_m\), and conjugate modal contributions combine to give real moments. The largest imaginary part of the retained spectrum is included among the diagnostics reported in Appendix~\ref{app:spectral-transient-implementation}.
The zero mode is the invariant cross-sectional density:
\begin{equation}
  \lambda_0=0,\qquad
  R_0=\rho_N,\qquad
  L_0=1.
\label{eq:polygon-zero-mode}
\end{equation}
The right zero mode is the long-time cross-sectional density, and the constant left zero mode represents conservation of mass. The non-zero right modes describe cross-sectional shapes left by the injection. A mode with a small \(\lambda_m\) persists longer and can influence the axial variance over a longer part of the transient.

The concentration is expanded as
\begin{equation}
  c(\x,z,t)=\sum_{m\ge0}a_m(z,t)R_m(\x),
  \qquad
  a_m(z,t)=\int_{\Omega_N}L_m(\x)c(\x,z,t)\,\dd A .
\label{eq:polygon-modal-expansion}
\end{equation}
The velocity profile couples the transverse modes through
\begin{equation}
  U_{mn}=
  \int_{\Omega_N}L_m(\x)u_N(\x)R_n(\x)\,\dd A .
\label{eq:polygon-velocity-matrix}
\end{equation}
Projecting \eqref{eq:polygon-leading-transient-model} gives the modal advection--relaxation system
\begin{equation}
  \frac{\partial a_m}{\partial t}
  +
  \Pe\sum_{n\ge0}U_{mn}\frac{\partial a_n}{\partial z}
  =
  -\lambda_m a_m .
\label{eq:polygon-modal-pde}
\end{equation}
The matrix \(U_{mn}\) is the Galerkin representation of multiplication by the Poiseuille velocity in the biorthogonal transverse basis. Its diagonal entries give the velocities sampled by individual relaxation structures, and its off-diagonal entries describe the advection-induced transfer of axial-gradient information between transverse modes. In Fourier space, with \(k\) the axial wavenumber and \(\widehat{\bm a}(k,t)\) the modal Fourier-amplitude vector,
\begin{equation}
  \frac{\dd\widehat{\bm a}}{\dd t}
  =
  -\left(\bm\Lambda+i k\Pe\,\bm U\right)\widehat{\bm a},
  \qquad
  \bm\Lambda=\operatorname{diag}(\lambda_0,\lambda_1,\ldots),
\label{eq:polygon-fourier-propagator}
\end{equation}
which is the propagator used for the field reconstructions in Figures~\ref{fig:transient-icA}--\ref{fig:transient-icC}.

The variance in Figure~\ref{fig:transient-running-kappa} is obtained from modal moments, avoiding an axial grid for the second moment. For
\(m_n^{(j)}(t)=\int z^j a_n(z,t)\,\dd z\), \(j=0,1,2\), integration by parts in \(z\) gives
\begin{equation}
  \frac{\dd}{\dd t}
  \begin{bmatrix}
    \bm m^{(0)}\\
    \bm m^{(1)}\\
    \bm m^{(2)}
  \end{bmatrix}
  =
  \begin{bmatrix}
    -\bm\Lambda & 0 & 0\\
    \Pe\bm U & -\bm\Lambda & 0\\
    0 & 2\Pe\bm U & -\bm\Lambda
  \end{bmatrix}
  \begin{bmatrix}
    \bm m^{(0)}\\
    \bm m^{(1)}\\
    \bm m^{(2)}
  \end{bmatrix}.
\label{eq:polygon-moment-system}
\end{equation}
The zero-mode components of this system give the packet mass, mean position and axial variance. The long-time Taylor coefficient has the modal representation
\begin{equation}
  \kappa_{{\rm TA},N}
  =
  \sum_{m\ge1}\frac{U_{0m}U_{m0}}{\lambda_m}.
\label{eq:polygon-modal-kappa}
\end{equation}
Equation~\eqref{eq:polygon-modal-kappa} is the modal Green-function representation of the cell problem in Section~\ref{subsec:taylor-coefficient} \citep{frankel1989foundations,ramirez2006,alexandre2021}. On the mass-conserving subspace, the inverse transverse relaxation operator is expanded in discrete form as \(\sum_{m\ge1}\lambda_m^{-1}\bm R_m\bm L_m^\ast\bm M\). The factor \(U_{0m}U_{m0}\) measures the coupling of mode \(m\) to the velocity deviation, while \(\lambda_m^{-1}\) measures how long that structure remains available to generate axial spreading. The computations retain 64 transverse modes. For the four polygons used below, the modal sum agrees with the cell-problem coefficient to relative differences between \(5.0\times10^{-5}\) and \(2.2\times10^{-4}\). The first relaxation eigenvalue increases from \(\lambda_1=1.436\) for \(N=3\) to \(\lambda_1=2.987\) for \(N=7\), while the Taylor coefficient decreases from \(1.2492\times10^{-2}\) to \(5.9423\times10^{-3}\), reflecting the simultaneous change in Poiseuille geometry and transverse relaxation.

\subsection{Cross-sectional relaxation from different initial conditions}
\label{subsec:cross-sectional-relaxation}

\begin{figure}[!htbp]
  \centering
  \includegraphics[width=0.92\linewidth]{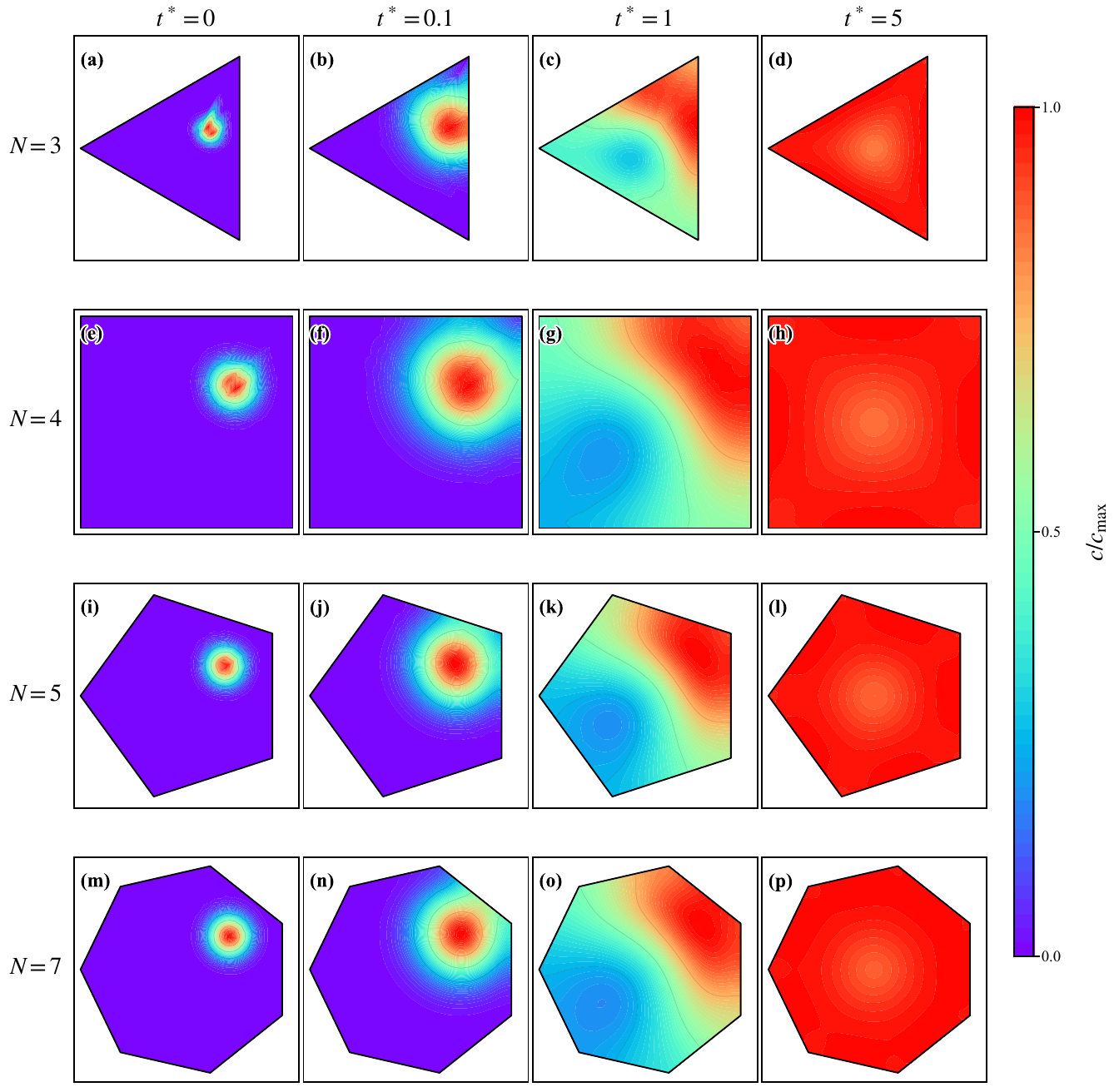}
  \caption{Transient cross-sectional relaxation for IC-A, a single off-centre pulse, at \(p=100\), \(\Pe=10^4\) and \(\Per=10\). Columns show \(t^\ast=0,0.1,1,5\), and rows show \(N=3,4,5,7\). The plotted field is \(c/c_{\max}\) at the moving pulse centre.}
  \label{fig:transient-icA}
\end{figure}

\begin{figure}[!htbp]
  \centering
  \includegraphics[width=0.92\linewidth]{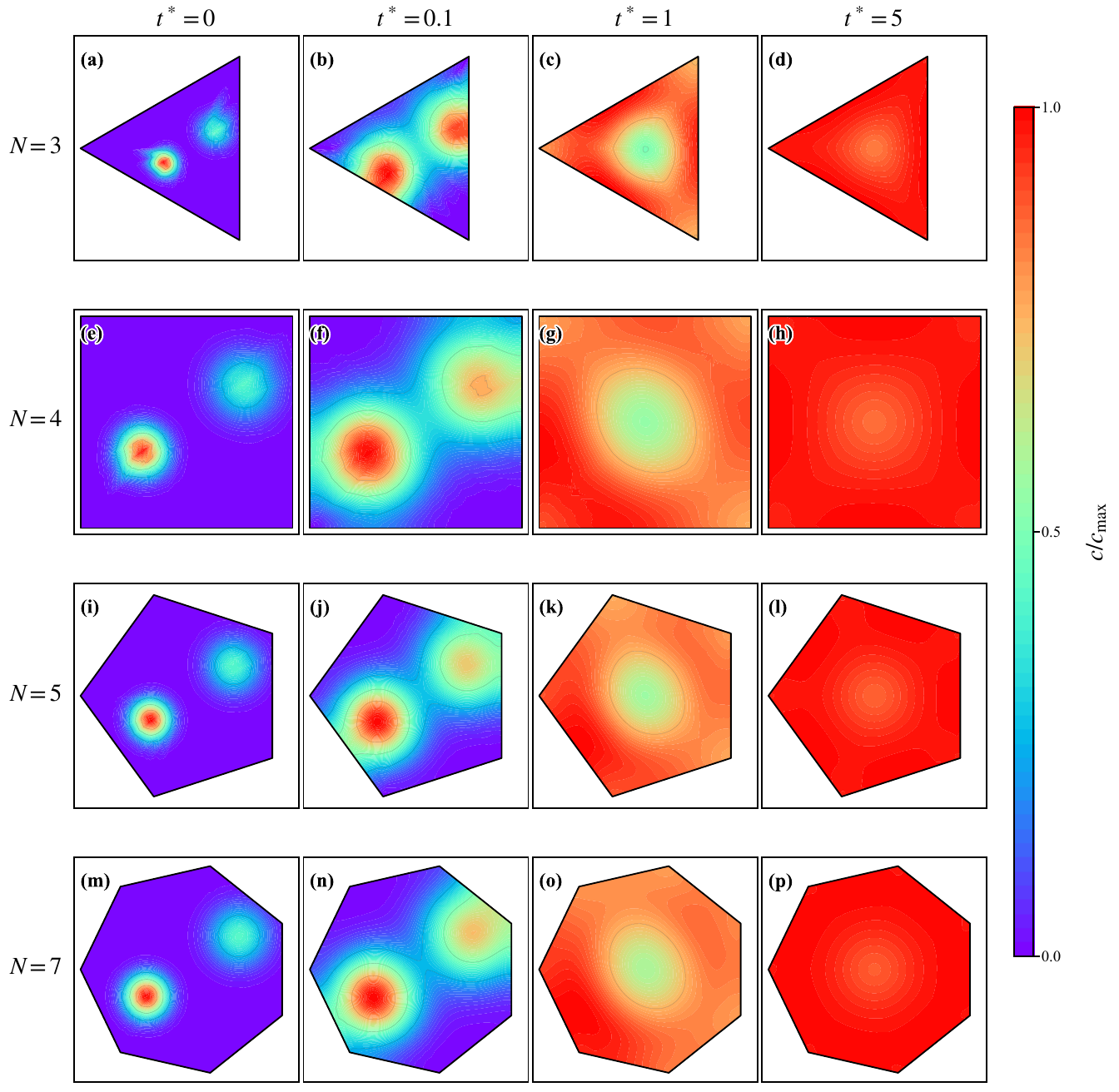}
  \caption{Transient cross-sectional relaxation for IC-B, an unequal two-pulse mixture, at the same parameter values as Figure~\ref{fig:transient-icA}. The layout and normalization are the same as in Figure~\ref{fig:transient-icA}.}
  \label{fig:transient-icB}
\end{figure}

\begin{figure}[!htbp]
  \centering
  \includegraphics[width=0.92\linewidth]{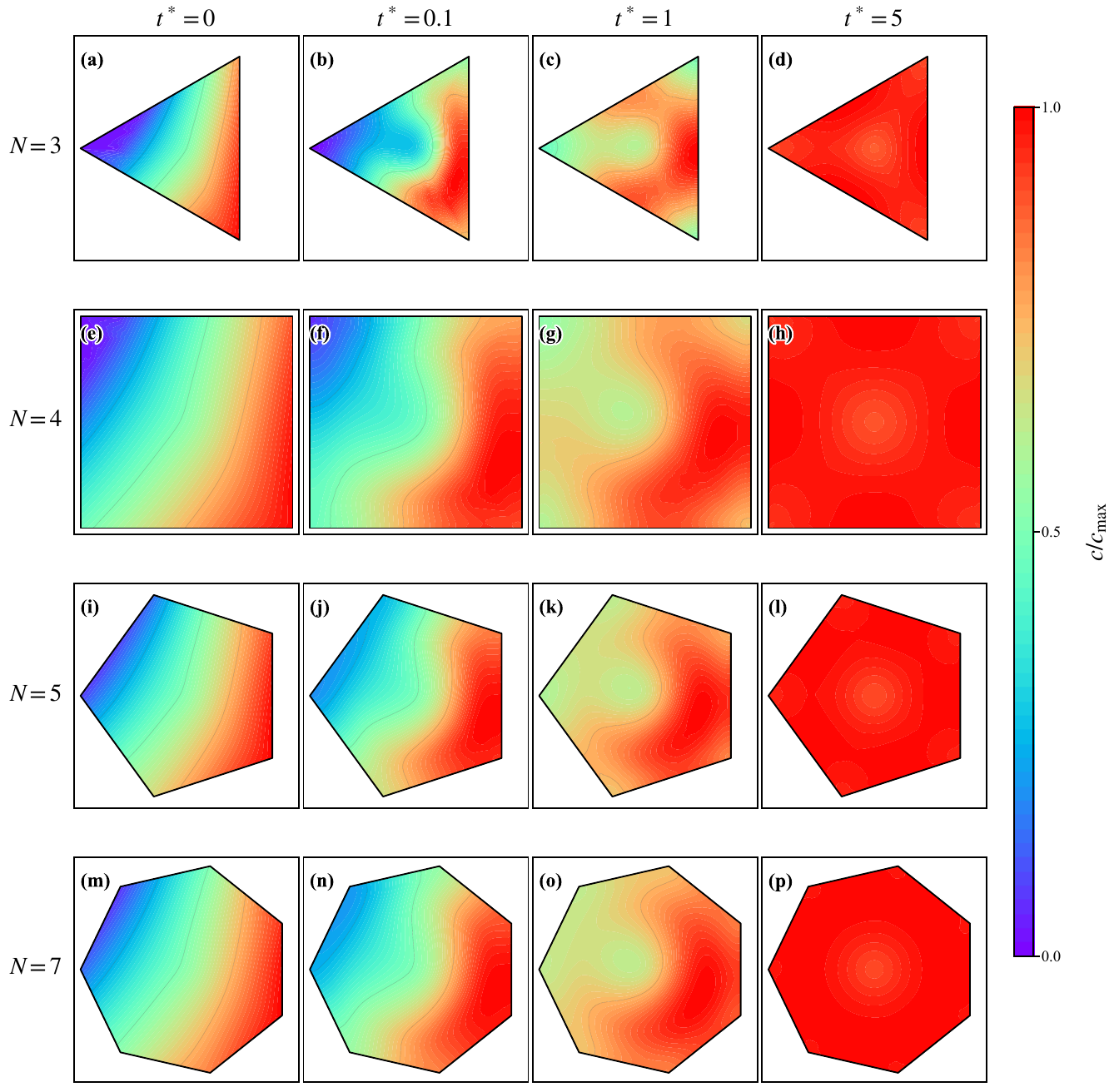}
  \caption{Transient cross-sectional relaxation for IC-C, a broad skewed profile weighted by the invariant density, at the same parameter values as Figure~\ref{fig:transient-icA}. The layout and normalization are the same as in Figure~\ref{fig:transient-icA}.}
  \label{fig:transient-icC}
\end{figure}

Figures~\ref{fig:transient-icA}--\ref{fig:transient-icC} show the same transient experiment for three different initial conditions (ICs). The axial part of the packet is the same in all cases, and only the transverse profile is varied. Writing \(\x=(x,y)\), define
\begin{equation}
  G_\sigma(\x;\x_0)=
  \exp\left[-\frac{|\x-\x_0|^2}{2\sigma^2}\right],
  \qquad
  \mathcal N_N[f]=
  \frac{f}{\int_{\Omega_N}f(\x)\,\dd A}.
\label{eq:transient-profile-normalisation}
\end{equation}
The three normalized transverse profiles are
\begin{subequations}
\label{eq:transient-initial-profiles}
\begin{align}
  g_{A,N}(\x)&=
  \mathcal N_N\!\left[
  G_{0.14}\!\left(\x;(0.45,0.35)\right)
  \right],
  \label{eq:transient-profile-A}\\
  g_{B,N}(\x)&=
  \mathcal N_N\!\left[
  0.70\,G_{0.15}\!\left(\x;(-0.42,-0.28)\right)
  +0.30\,G_{0.21}\!\left(\x;(0.55,0.35)\right)
  \right],
  \label{eq:transient-profile-B}\\
  g_{C,N}(\x)&=
  \mathcal N_N\!\left[
  \rho_N(\x)\max\{0.05,\,1+0.70x-0.32y+0.22xy\}
  \right].
  \label{eq:transient-profile-C}
\end{align}
\end{subequations}
Thus IC-A is a localized off-centre pulse, IC-B is an unequal two-pulse mixture, and IC-C is a broad skewed perturbation of the invariant density. The three profiles probe different parts of the transverse spectrum. IC-A is localized and contains substantial high-mode content. IC-B tests a multi-region injection that initially samples separated streamline velocities. IC-C starts close to \(\rho_N\) and mainly probes the decay of low-mode skewness. The full initial condition is \(c(\x,z,0)=g_{J,N}(\x)h(z)\), \(J=A,B,C\), with the same normalized axial Gaussian \(h(z)\propto\exp[-(z-z_0)^2/(2\sigma_z^2)]\) in all three cases. The plotted cross-section is taken at the moving pulse centre \(z_c(t)=z_0+\Pe\bar u_Nt\), so the panels follow transverse relaxation after subtracting the bulk axial translation. Each panel is normalized by its own maximum concentration.

\FloatBarrier

IC-A gives the most localized transverse injection among the three cases. At \(t^\ast=0\), most of the mass is concentrated away from the centre and away from the wall. By \(t^\ast=0.1\), modes with large \(\lambda_m\) have decayed enough to broaden the peak, while lower modes still retain an off-centre asymmetry. By \(t^\ast=1\), the residual field is mainly controlled by the lowest non-zero modes, whose shapes are set by the polygonal boundary and the shear-dependent mobility tensor. The \(t^\ast=5\) column is close to the zero-mode invariant density for every \(N\); the remaining contrast is a weak near-wall enrichment, about \(2\)--\(4\%\) above the cross-sectional mean for this initial condition.

\FloatBarrier

IC-B starts with two separated concentration regions and therefore excites a different combination of non-zero transverse modes. The unequal weights make the early cross-section sample two velocity regions at once, giving finite amplitudes both in localized high modes and in lower modes representing the contrast between the two regions. The separated peaks merge as the high modes decay, and the field at \(t^\ast=O(1)\) is dominated by the slower residual structure. The triangular case keeps the strongest geometric imprint, while \(N=5\) and \(N=7\) already display a more circular-like relaxation pattern by \(t^\ast=1\). The late-time field converges to the zero right mode \(\rho_N\), with near-wall enrichment comparable to IC-A.

\FloatBarrier

IC-C begins much closer to the invariant structure because its broad envelope is weighted by \(\rho_N\). Its early field still contains a skewed transverse memory, visible most clearly in the \(t^\ast=0\) and \(t^\ast=0.1\) columns. The weaker high-mode content gives a smaller early-time shape change than in IC-A and IC-B, and the late-time near-wall contrast is weaker. Taken together, Figures~\ref{fig:transient-icA}--\ref{fig:transient-icC} show that localized, multi-peaked and broad invariant-weighted injections all lose their non-zero modal content and converge to the same invariant sampling state.

\subsection{From transient memory to the asymptotic Taylor--Aris regime}
\label{subsec:running-diffusivity}

\begin{figure}[!htbp]
  \centering
  \includegraphics[width=0.92\linewidth]{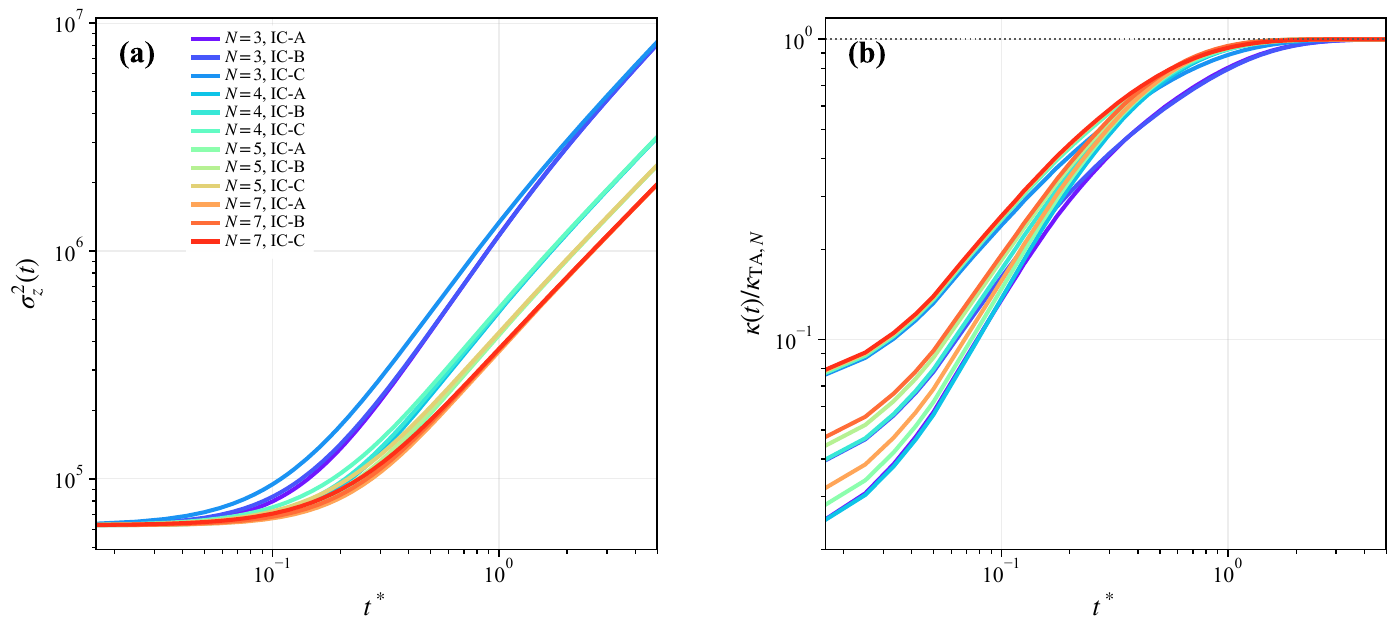}
  \caption{Transient variance growth and running Taylor coefficient for the cases in Figures~\ref{fig:transient-icA}--\ref{fig:transient-icC}. Panel (a) shows the axial variance \(\sigma_z^2(t)\) against \(t^\ast\). Panel (b) shows the running coefficient normalized by the case-specific long-time Taylor value \(\kappa_{{\rm TA},N}\).}
  \label{fig:transient-running-kappa}
\end{figure}

The cross-sectional relaxation in Figures~\ref{fig:transient-icA}--\ref{fig:transient-icC} has a direct axial-moment consequence. Figure~\ref{fig:transient-running-kappa} computes the axial variance from the modal moment system \eqref{eq:polygon-moment-system} and compares its finite-time growth rate with the cell-problem value. The comparison targets the leading high-\(\Pe\) Taylor contribution, with the lower-order direct-diffusion and cross-drift corrections treated as described after \eqref{eq:polygon-leading-transient-model}. In the Taylor regime \citep{taylor1953,aris1956,vedel2012,vedel2014},
\begin{equation}
  \sigma_z^2(t)
  \sim
  2\Pe^2\kappa_{{\rm TA},N}t
  \qquad (t\to\infty),
\label{eq:transient-variance-growth}
\end{equation}
so we define the running Taylor coefficient by
\begin{equation}
  \kappa(t)
  =
  \frac{1}{2\Pe^2}
  \frac{\dd\sigma_z^2}{\dd t},
\label{eq:transient-running-kappa}
\end{equation}
which approaches \(\kappa_{{\rm TA},N}\) when the packet reaches the Taylor--Aris regime.

\FloatBarrier

Panel (a) shows the variance entering a regime in which \(d\sigma_z^2/dt\) is independent of the initial transverse profile for each polygon. The early growth rates differ because IC-A, IC-B and IC-C initially project onto different non-zero transverse modes. These modes bias the packet toward different parts of the Poiseuille profile and therefore produce different finite-time variance growth. After the velocity-coupled modes decay over \(t^\ast=O(1)\), the leading variance growth approaches \(2\Pe^2\kappa_{{\rm TA},N}\).

Panel (b) shows the same convergence through the running coefficient. All twelve curves approach \(\kappa(t)/\kappa_{{\rm TA},N}=1\). At the final time \(t^\ast=5\), the ratios lie within about \(6.7\times10^{-4}\) of unity for \(N=3\), and within about \(2.4\times10^{-5}\) of unity for \(N=4\); the \(N=5\) and \(N=7\) cases are within \(1.3\times10^{-5}\). The mass drift in the moment calculation remains at the level of \(10^{-11}\) or smaller. Thus the coefficient obtained from the cell problem is the long-time dispersion coefficient selected by finite injections under the leading high-\(\Pe\) dynamics.

Figures~\ref{fig:transient-icA}--\ref{fig:transient-icC} identify the cross-sectional mechanism: different injections excite different non-zero transverse modes, and those modes decay on the transverse relaxation scale until the zero-mode invariant density remains. Figure~\ref{fig:transient-running-kappa} gives the axial-moment consequence: after this modal memory has decayed, the variance growth rate is independent of the injection protocol and equals the cell-problem value \(\kappa_{{\rm TA},N}\). The spectral transient calculation therefore establishes \(\kappa_{{\rm TA},N}\) as the long-time variance-growth coefficient selected by the leading high-\(\Pe\) polygonal rod dynamics.

\section{Conclusions}
\label{sec:conclusions}

We have developed a Taylor--Aris theory for dilute Brownian rods in pressure-driven flow through regular-polygonal ducts. The formulation uses a local shear-aligned frame to combine the Jeffery--Brownian orientational equilibrium with the cross-sectional Poiseuille field. This separates the local rod response, which depends on the local rotational P{\'e}clet number \(q_N(\x;\Per)\) and aspect ratio \(p\), from the polygonal geometry, which determines how the local transport coefficients are distributed in the section.

The orientation closure shows that shear alignment reduces transverse mobility, enhances axial diffusion and produces a signed cross coefficient associated with the mixed shear-plane moment. When these coefficients are placed in the polygonal section, the transverse relaxation operator selects an invariant density \(\rho_N\), rather than the area measure. In the circular limit this reduces to the explicit weighting \(\rho_\infty\propto D_s^{-1}\); for finite polygons it remains a two-dimensional problem involving \(D_s\), \(D_\eta\) and the spatially varying shear direction.

The reduced one-dimensional model contains a leading speed \(\Pe\bar u_N\), a leading Taylor diffusivity \(\Pe^2\kappa_N\), and lower-order contributions from direct axial diffusion and from the conservative drift induced by the signed shear--axial cross coefficient \(A\). The steady calculations show that alignment has a small, non-monotone effect on \(\bar u_N\), because the invariant density shifts the streamline sampling. The effect on \(\kappa_N\) is larger because reduced transverse relaxation increases the Taylor cell response. The normalized enhancement approaches the fully aligned transverse-mixing scale as \(\Per\) increases.

The finite-polygon results converge smoothly to the circular-pipe branch. Small polygons, especially the triangular duct, retain distinct shear distributions and cell responses, whereas \(N\gtrsim12\) gives a close approximation to the circular result for the steady coefficients considered here.

The spectral calculation gives the corresponding finite-time route to this asymptotic state. It resolves the transverse relaxation modes of the same operator that defines \(\rho_N\) and \(\kappa_N\). The zero mode is the invariant density, while the non-zero modes carry the memory of the injection profile. Localized, multi-peaked and broad initial distributions therefore produce different pre-asymptotic variance growth before transverse equilibration. As these modes decay, the instantaneous Taylor coefficient \(\kappa(t)=(2\Pe^2)^{-1}\mathrm d\sigma_z^2/\mathrm dt\) becomes independent of the injection protocol and converges to the cell-problem value \(\kappa_{{\rm TA},N}\). This identifies how finite injections select the asymptotic Taylor--Aris coefficient.

These results also indicate several natural extensions. The point-particle approximation could be relaxed to incorporate finite-size wall effects, which may promote localized rod trapping or depletion near the sharp corners of low-\(N\) polygons. Extending the present non-radial geometric framework to semi-dilute suspensions or active microswimmers would clarify how inter-particle interactions and self-propulsion compete with the polygonal shear field. Finally, coupling the local shear-aligned formulation to shape-optimization protocols offers a route to the inverse design of microfluidic channels with tailored transport dispersivity.

\appendix
\setcounter{section}{0}
\setcounter{figure}{0}
\setcounter{table}{0}
\renewcommand{\thesection}{\Alph{section}}
\renewcommand{\theequation}{\thesection.\arabic{equation}}
\renewcommand{\thefigure}{\thesection\arabic{figure}}
\renewcommand{\thetable}{\thesection\arabic{table}}
\makeatletter
\@addtoreset{figure}{section}
\@addtoreset{table}{section}
\renewcommand{\theHsection}{appendix.\Alph{section}}
\renewcommand{\theHequation}{appendix.\Alph{section}.\arabic{equation}}
\renewcommand{\theHfigure}{appendix.\Alph{section}.\arabic{figure}}
\renewcommand{\theHtable}{appendix.\Alph{section}.\arabic{table}}
\makeatother

\section{Local orientation solver and closure diagnostics}
\label{app:local-orientation-solver}

The local closure in Section~\ref{subsec:local-closure} is evaluated as a function of the scalar shear strength \(q\) and the aspect ratio \(p\). This appendix records the numerical angular problem used to compute the closure functions \(D_s(q;p)\), \(D_\eta(q;p)\), \(B(q;p)\) and \(A(q;p)\). The calculation is independent of the polygonal cross-section; the polygon only supplies the local value of \(q_N(\x;\Per)\).

In the local shear frame, write
\begin{equation}
  p_s=(1-\mu^2)^{1/2}\sin\theta,
  \qquad
  p_\eta=\mu,
  \qquad
  p_z=(1-\mu^2)^{1/2}\cos\theta ,
\label{eq:app-orientation-coordinates}
\end{equation}
with \(0\le\theta<2\pi\) and \(-1\le\mu\le1\). Here \(\mu=\sin\psi\), with \(\psi\) the inclination out of the local \(s\)-\(z\) shear plane, and \(\beta=(p^2-1)/(p^2+1)\) is the Jeffery shape factor used below. The steady Jeffery--Brownian balance is solved in the dimensionless form
\begin{equation}
  \Delta_p g
  +
  2q\,\mathcal J_\beta g=0,
  \qquad
  \int_{S^2}g\,\dd\Omega=1 ,
\label{eq:app-angular-pde}
\end{equation}
where \(S^2\) is the unit orientation sphere, \(\dd\Omega\) is its surface element, \(\pvec\in S^2\), and \(\Delta_p\) is the surface Laplacian. The angular drift operator is
\begin{equation}
  \mathcal J_\beta g
  =
  \partial_\theta(\omega_\theta g)
  +
  \partial_\mu(\omega_\mu g),
\label{eq:app-jeffery-operator}
\end{equation}
with
\begin{equation}
  \omega_\theta=\frac12(1-\beta\cos2\theta),
  \qquad
  \omega_\mu=\frac12\beta\sin2\theta\,\mu(1-\mu^2).
\label{eq:app-jeffery-drift}
\end{equation}
This is the angular equation implemented by the preprocessing solver. For \(\beta=0\) the drift is divergence-free with respect to the isotropic density and \(g=(4\pi)^{-1}\).

The density is represented by even real spherical harmonics,
\begin{equation}
  g_L(\theta,\mu;q)
  =
  \frac1{4\pi}
  +
  \sum_{\ell=1}^{L_o}a_{\ell0}(q)Y_{2\ell,0}
  +
  \sum_{\ell=1}^{L_o}\sum_{m=1}^{\ell}
  \left[
  a_{\ell m}^c(q)Y_{2\ell,2m}^c
  +
  a_{\ell m}^s(q)Y_{2\ell,2m}^s
  \right].
\label{eq:app-angular-expansion}
\end{equation}
Here \(Y_{2\ell,0}\), \(Y_{2\ell,2m}^c\) and \(Y_{2\ell,2m}^s\) are real spherical harmonics even under \(\pvec\mapsto-\pvec\), \(L_o\) is the angular truncation, and the coefficients \(a_{\ell0}\), \(a_{\ell m}^c\) and \(a_{\ell m}^s\) are collected in the vector \(\bm a(q)\). The constant mode is not included among the unknowns; it is fixed by the normalization. Galerkin projection gives, at each sampled value of \(q\),
\begin{equation}
  \left(\bm L+2q\,\bm J_\beta\right)\bm a(q)
  =
  -2q\,\bm b_\beta ,
\label{eq:app-angular-linear-system}
\end{equation}
where \(\bm L\) is the projection of \(\Delta_p\), \(\bm J_\beta\) is the projection of \(\mathcal J_\beta\) on the non-constant basis functions, and \(\bm b_\beta\) is the projection of \(\mathcal J_\beta(4\pi)^{-1}\). The quadrature uses a uniform rule in \(\theta\) and Gauss--Legendre points in \(\mu\). For the results reported here, the closure calculation uses the same harmonic truncation and quadrature as those used to generate the figures; lower truncations are used only for the built-in self-tests.

\begin{figure}
  \centering
  \includegraphics[width=0.92\linewidth]{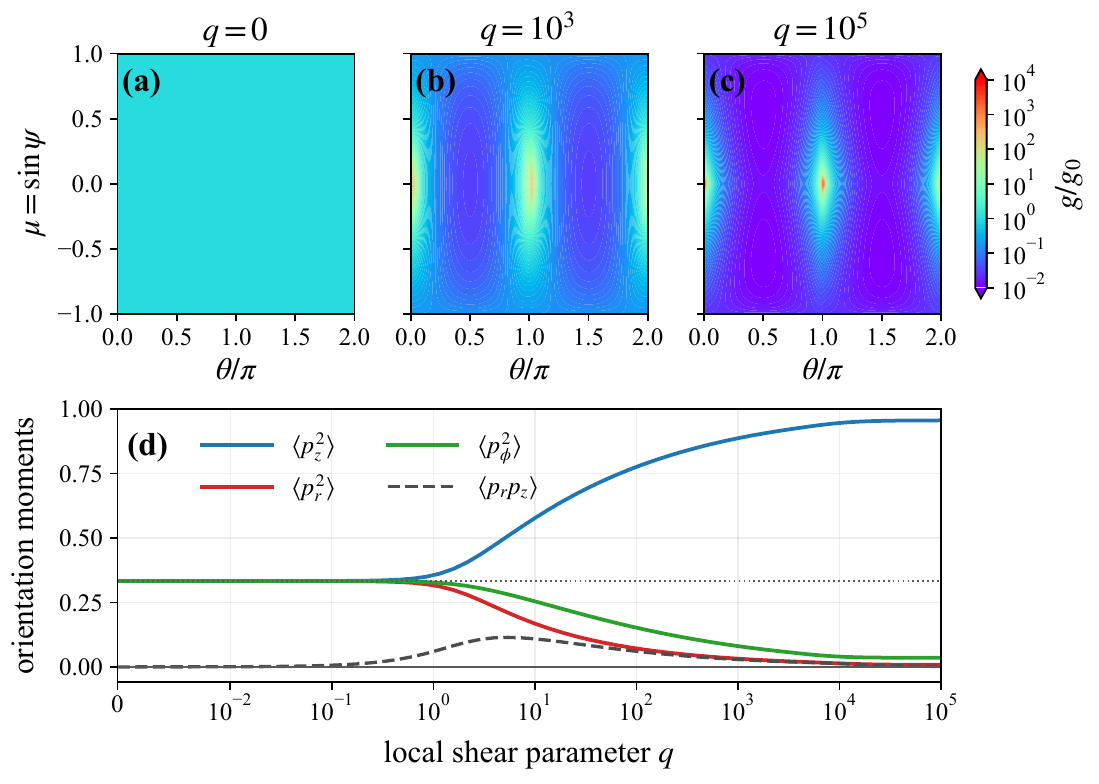}
  \caption{Representative angular distribution and orientation moments used by the local closure. The example uses \(p=1000\), corresponding to \(\beta=0.999998\), \(D_\parallel/\Dbar=1.4013\) and \(D_\perp/\Dbar=0.7993\). Panels (a)--(c) show the probability density \(g/g_0\) in \((\theta,\mu)\) coordinates at \(q=0\), \(q=10^3\) and \(q=10^5\), where \(\mu=\sin\psi\) and \(g_0=(4\pi)^{-1}\). Panel (d) shows the second-order moments \(\avg{p_z^2}\), \(\avg{p_r^2}\), \(\avg{p_\phi^2}\) and \(\avg{p_rp_z}\) as functions of \(q\). In the polygonal notation used in the main text, \(p_r\) and \(p_\phi\) correspond to \(p_s\) and \(p_\eta\), respectively.}
  \label{fig:app-orientation-density-moments}
\end{figure}

After \(g_L\) is reconstructed and renormalized at each sampled \(q\), the four closure functions are computed from the moments in \eqref{eq:local-functions}. The \(q\)-grid is linear near \(q=0\) and logarithmic at larger \(q\), and monotone piecewise-cubic interpolation is used when evaluating the closure functions on the polygonal mesh. The sampled range is chosen to cover the largest value of \(q_N\) used in the cross-section.

The following diagnostics are recorded for each closure calculation:
\begin{subequations}
\label{eq:app-closure-diagnostics}
\begin{align}
  \epsilon_{\rm norm}
  &=
  \max_q\left|\int_{S^2}g_L\,\dd\Omega-1\right|,\\
  g_{\min}
  &=
  \min_{q,\pvec}g_L(\pvec;q),\\
  \epsilon_{\rm tr}
  &=
  \max_q |D_s(q)+D_\eta(q)+B(q)-3|,\\
  \delta_{\rm sz}
  &=
  \min_q \{D_s(q)B(q)-A^2(q)\}.
\end{align}
\end{subequations}
Here \(d_\parallel=D_\parallel/\Dbar\) and \(d_\perp=D_\perp/\Dbar\). The trace diagnostic checks the normalization \((d_\parallel+2d_\perp)/3=1\), and \(\delta_{\rm sz}>0\) checks positive definiteness of the local \(s\)-\(z\) diffusion block. The spherical branch is handled analytically in the code, giving \(D_s=D_\eta=B=1\) and \(A=0\) for every \(q\).

\section{Finite-element discretisation and steady solver diagnostics}
\label{app:fem-steady-solver}

The finite-polygon computations use continuous piecewise-linear finite elements on a polar-star triangulation of \(\Omega_N\). Let
\begin{equation}
  \theta_\ell=\frac{2\pi\ell}{n_\theta},
  \qquad
  n_\theta=2Nn_f,
\end{equation}
where \(n_f\) is the angular refinement factor. The boundary radius in direction \(\theta\) is
\begin{equation}
  R_N(\theta)
  =
  \left[
  \max_{0\le j<N}
  \cos\left(\theta-\frac{2\pi j}{N}\right)
  \right]^{-1}.
\label{eq:app-polygon-radius}
\end{equation}
The mesh nodes are
\begin{equation}
  \x_{i\ell}
  =
  \frac{i}{n_r}R_N(\theta_\ell)
  (\cos\theta_\ell,\sin\theta_\ell),
  \qquad
  i=0,\ldots,n_r ,
\label{eq:app-polygon-mesh}
\end{equation}
Here \(n_r\) is the number of radial intervals, with the \(i=0\) ring collapsed to the centre. Consecutive rings are split into triangles. Since the angular grid contains both side-normal and vertex directions, the outer polygon is represented exactly up to round-off. All sums over \(T\) below run over these triangles, and \(|T|\) denotes the triangle area.

Let \(V_h\) be the continuous \(P_1\) finite-element space on this triangulation, let \(\{\phi_i\}\) be its nodal basis, and let \(V_h^0\subset V_h\) denote the subspace with zero boundary values. The stiffness matrix used below is
\[
  K_{ij}=\int_{\Omega_N}\nabla\phi_i\cdot\nabla\phi_j\,\dd A .
\]
The Poiseuille solve uses the weak form
\begin{equation}
  \int_{\Omega_N}\nabla v_h\cdot\nabla\tilde u_h\,\dd A
  =
  \int_{\Omega_N}v_h\,\dd A
  \qquad
  \text{for all }v_h\in V_h^0 ,
\label{eq:app-poiseuille-weak}
\end{equation}
with \(\tilde u_h=0\) on boundary nodes. The normalized velocity is \(u_h=\tilde u_h/\max\tilde u_h\). Elementwise velocity gradients give \(\e_s\) and \(\e_\eta\); nodal area-averaged gradients are used to evaluate \(q_N\) and then the local closure functions.

Let \(M\) be the consistent mass matrix, \(M_{ij}=\int_{\Omega_N}\phi_i\phi_j\,\dd A\), let \(m_i=\int_{\Omega_N}\phi_i\,\dd A\), let \(\bm m=(m_i)\), and let \(\bm 1\) denote the vector of ones of compatible length. The vector \(\bm u\) contains the nodal values of \(u_h\). The matrix \(A_0\) represents the positive relaxation operator \(-\mathcal L_{0,N}\). With \(D_s^h\) and \(D_\eta^h\) the nodal \(P_1\) interpolants and with \(\e_s,\e_\eta\) constant on each element,
\begin{equation}
  (A_0)_{ij}
  =
  \sum_T
  \int_T
  \nabla\phi_i\cdot
  \left[
  \e_s\e_s^T\nabla(D_s^h\phi_j)
  +
  \e_\eta\e_\eta^T\nabla(D_\eta^h\phi_j)
  \right]\,\dd A .
\label{eq:app-A0-assembly}
\end{equation}
A three-point barycentric quadrature is used for the product terms. This assembly differentiates the conservative products \(D_s c\) and \(D_\eta c\), matching \eqref{eq:L0N}. Conservation is checked by the column-sum diagnostic
\begin{equation}
  \epsilon_{\rm col}
  =
  \|\bm 1^T A_0\|_\infty .
\label{eq:app-A0-colsum}
\end{equation}

The invariant density \(\rho_h=\sum_i\rho_i\phi_i\), with coefficient vector \(\bm\rho=(\rho_i)\), is obtained from the constrained saddle-point system
\begin{equation}
  \begin{bmatrix}
    A_0 & \bm m\\
    \bm m^T & 0
  \end{bmatrix}
  \begin{bmatrix}
    \bm\rho\\
    \lambda
  \end{bmatrix}
  =
  \begin{bmatrix}
    \bm 0\\
    1
  \end{bmatrix}.
\label{eq:app-rho-saddle}
\end{equation}
where \(\lambda\) is the Lagrange multiplier for the mass constraint. The numerical mass is then renormalized to \(\bm m^T\bm\rho=1\). The mean velocity is
\begin{equation}
  \bar u_N^h=\bm u^TM\bm\rho .
\label{eq:app-ubar-discrete}
\end{equation}
For the cell problem, define
\begin{equation}
  f_i=\int_{\Omega_N}\phi_i(u_h-\bar u_N^h)\rho_h\,\dd A .
\label{eq:app-cell-load}
\end{equation}
and collect these entries in \(\bm f=(f_i)\). Because \(A_0\) represents \(-\mathcal L_{0,N}\), the discrete corrector \(\chi_h=\sum_i\chi_i\phi_i\), with coefficient vector \(\bm\chi=(\chi_i)\), satisfies
\begin{equation}
  \begin{bmatrix}
    A_0 & \bm m\\
    \bm m^T & 0
  \end{bmatrix}
  \begin{bmatrix}
    \bm\chi\\
    \lambda_\chi
  \end{bmatrix}
  =
  \begin{bmatrix}
    -\bm f\\
    0
  \end{bmatrix}.
\label{eq:app-chi-saddle}
\end{equation}
where \(\lambda_\chi\) is the Lagrange multiplier enforcing the zero-mass constraint.
The Taylor coefficient is evaluated as
\begin{equation}
  \kappa_N^h
  =
  -(\bm u-\bar u_N^h\bm 1)^TM\bm\chi .
\label{eq:app-kappa-discrete}
\end{equation}
The cross-diffusive drift is computed elementwise from the same \(P_1\) fields,
\begin{equation}
  U_{A,N}^h
  =
  -\sum_T |T|\,
  \e_{s,T}\cdot
  \nabla(A^h\rho^h)\big|_T .
\label{eq:app-uA-discrete}
\end{equation}
where \(A^h\) and \(\rho^h\) denote the nodal \(P_1\) interpolants of \(A\) and \(\rho_h\).

The steady solver records the pre-renormalization mass error, \(\epsilon_{\rm col}\), the residuals
\begin{equation}
  \epsilon_\rho
  =
  \frac{\|A_0\bm\rho\|_2}{\max(1,\|\bm\rho\|_2)},
  \qquad
  \epsilon_\chi
  =
  \frac{\|A_0\bm\chi+\bm f\|_2}{\max(1,\|\bm f\|_2)},
\label{eq:app-steady-residuals}
\end{equation}
and the compatibility defect \(\sum_i f_i\). The built-in checks include exact polygon area recovery, positivity of element areas, zero boundary velocity, the identity \(A_0=K\) for constant scalar diffusion, the scalar variable-diffusion null mode \(\rho\propto D^{-1}\), the spherical branch \(R_{\kappa,N}=\kappa_N/\kappa_{s,N}=1\), and \(U_{A,N}=0\) for \(p=1\) and for \(\Per=0\).

The steady calculations reported here use the maximum-shear normalization in \eqref{eq:qN-definition}. The main \(\Per\)-sweeps in Figures~\ref{fig:mean-speed-polygons}--\ref{fig:enhancement-collapse-polygons} use \(n_r=64,n_f=12\) for \(N=3\) and \(n_r=48,n_f=10\) for \(N=4,5,6,8,12\); the field maps in Figures~\ref{fig:fields-p10}--\ref{fig:fields-p1000} use \(n_r=64,n_f=12\). The polygon-to-pipe convergence calculation in Figure~\ref{fig:polygon-pipe-convergence} uses finer meshes, with \(n_r=80,n_f=16\) for \(N=3\), \(n_r=64,n_f=12\) for \(N=4,5,6,8\), and \(n_r=96,n_f=20\) for \(N=12,16,24\). In the high-shear mesh-convergence check \(p=\infty\), \(\Per=10^4\), the change in \(\kappa_N\) between the two finest Taylor grids is \(5.2\times10^{-3}\) for \(N=3\), at most \(2.1\times10^{-3}\) for \(N=4\), and below \(1.1\times10^{-3}\) for \(N\ge5\).

\section{Spectral transient implementation}
\label{app:spectral-transient-implementation}

The transient figures use the same \(A_0\), \(M\), \(u_h\), \(D_s\), \(D_\eta\), \(B\) and \(\rho_h\) as the steady solver. The left and right generalized eigenvectors are computed from \eqref{eq:polygon-spectral-eigenproblem} and normalized as in \eqref{eq:polygon-biorthogonality}; the asterisk denotes conjugate transpose. Dense eigensolves are used for the meshes in Figures~\ref{fig:transient-icA}--\ref{fig:transient-running-kappa}; a shift-invert sparse option is used for larger convergence checks. The zero right eigenvector is mass-normalized and compared with the constrained solve for \(\rho_h\), while the zero left eigenvector is compared with the constant vector.

The transient calculations reported in Figures~\ref{fig:transient-icA}--\ref{fig:transient-running-kappa} use \(n_r=24,n_f=4\) and retain 64 transverse modes for each polygon. For the plotted cases, the truncated modal sum for \(\kappa_{{\rm TA},N}\) agrees with the corresponding cell-problem value to relative errors below \(2.2\times10^{-4}\).

The velocity and direct-diffusion matrices are
\begin{equation}
  M_u=(\int_{\Omega_N}\phi_i u_h\phi_j\,\dd A)_{ij},
  \qquad
  M_B=(\int_{\Omega_N}\phi_i B_h\phi_j\,\dd A)_{ij}.
\label{eq:app-weighted-mass}
\end{equation}
Projection gives
\begin{equation}
  B_{mn}^{\rm dir}=\bm L_m^\ast M_B\bm R_n .
\label{eq:app-modal-matrices}
\end{equation}
The velocity matrix \(U_{mn}\) is the discrete weighted-mass evaluation of \eqref{eq:polygon-velocity-matrix}. Here \(B_h\) is the nodal interpolant of the direct axial diffusivity \(B(\x)\), and \(\bm B^{\rm dir}=(B_{mn}^{\rm dir})\). For a separable initial condition \(c(\x,z,0)=c_\perp^0(\x)h(z)\), the Fourier-modal coefficients are initialized by
\begin{equation}
  \widehat a_m(k,0)
  =
  \bm L_m^\ast M\bm c_\perp^0\,\widehat h(k).
\label{eq:app-modal-initial-projection}
\end{equation}
Here \(\bm c_\perp^0\) contains the nodal values of the transverse injection profile, and \(\widehat h(k)\) is the Fourier coefficient of the axial pulse.
The reconstructed fields in Figures~\ref{fig:transient-icA}--\ref{fig:transient-icC} use
\begin{equation}
  \widehat{\bm a}(k,t)
  =
  \exp[-(\bm\Lambda+i k\Pe\,\bm U)t]\,
  \widehat{\bm a}(k,0),
\label{eq:app-modal-fourier-evolution}
\end{equation}
where \(\bm\Lambda=\operatorname{diag}(\lambda_m)\). The optional \(k^2\bm B^{\rm dir}\) term is retained only in diagnostic runs; the plotted Taylor-regime relaxation isolates the leading high-\(\Pe\) advection--relaxation balance.

The transverse profiles for Figures~\ref{fig:transient-icA}--\ref{fig:transient-icC} are formed on the FEM nodes and normalized with the lumped mass vector \(\bm m\). Negative values are clipped before normalization only for robustness of the broad skewed profile. The axial pulse is a periodic Gaussian on a domain of length \(1.2\Pe\), with standard deviation \(0.025\Pe\). The figure scripts retain Fourier modes whose initial amplitudes exceed a relative tolerance and sample the cross-section at the moving centre \(z_0+\Pe\bar u_Nt\), where \(z_0\) is the initial axial centre.

For Figure~\ref{fig:transient-running-kappa}, the moment system \eqref{eq:polygon-moment-system} is integrated directly by a matrix-exponential action. For each initial profile, the nodal zeroth, first and second axial moments are reconstructed from the retained modes and integrated with the mass vector. Let \(\bm c^{(j)}(t)\) be the nodal vector with entries \(\int z^j c(\x_i,z,t)\,\dd z\), for \(j=0,1,2\). The variance is
\begin{equation}
  \sigma_z^2(t)
  =
  \frac{\bm m^T\bm c^{(2)}(t)}{\bm m^T\bm c^{(0)}(t)}
  -
  \left[
  \frac{\bm m^T\bm c^{(1)}(t)}{\bm m^T\bm c^{(0)}(t)}
  \right]^2 .
\label{eq:app-variance-discrete}
\end{equation}
The running coefficient follows the definition in \eqref{eq:transient-running-kappa}. The output diagnostics include \(|\lambda_0|\), the largest imaginary part of the retained relaxation eigenvalues, biorthogonality error, zero-mode errors, modal-to-cell \(\kappa\) error, final mass drift and the relative imaginary residual in reconstructed real fields.

\section{Full transverse-space Fourier--FEM small-wavenumber validation}
\label{app:full-fourier-fem-validation}

The full transverse-space Fourier--FEM calculation is used as an independent small-\(k\) validation of the reduced spectral model in Section~\ref{sec:transient-spectral-validation}. It works in the complete transverse finite-element space and does not use a transverse modal truncation. For a single axial Fourier mode,
\begin{equation}
  c(\x,z,t)=\widehat c_k(\x,t)e^{ikz},
\end{equation}
the reference implementation solves
\begin{equation}
  M\frac{\dd\widehat{\bm c}_k}{\dd t}
  =
  -H(k)\widehat{\bm c}_k,
  \qquad
  H(k)=A_0+i k\Pe\,M_u+k^2M_B .
\label{eq:app-full-fourier-operator}
\end{equation}
Here \(\widehat{\bm c}_k\) is the nodal coefficient vector of \(\widehat c_k\), and \(M\), \(A_0\), \(M_u\) and \(M_B\) are the finite-element matrices defined in Appendices~\ref{app:fem-steady-solver} and \ref{app:spectral-transient-implementation}. The \(k^2M_B\) term is the full-space counterpart of direct axial diffusion.
The principal generalized eigenvalue is defined by
\begin{equation}
  H(k)\bm r(k)=\alpha(k)M\bm r(k),
\label{eq:app-full-fourier-eigenproblem}
\end{equation}
where \(\bm r(k)\) is the corresponding generalized right eigenvector, with the branch selected near \(i k\Pe\bar u_N\). As \(k\to0\),
\begin{equation}
  \alpha(k)
  =
  i k\Pe\,\bar u_N
  +
  k^2\left(K_{{\rm dir},N}+\Pe^2\kappa_N\right)
  +
  O(k^3).
\label{eq:app-small-k-expansion}
\end{equation}
Here \(K_{{\rm dir},N}\) is the direct axial diffusivity defined in \eqref{eq:KdirN-definition}; in the same finite-element discretization it is evaluated as \(\bm 1^TM_B\bm\rho\).
Thus each non-zero sample gives
\begin{subequations}
\label{eq:app-small-k-extraction}
\begin{align}
  \bar u_N^{\rm full}(k)
  &=
  \frac{\operatorname{Im}\alpha(k)}{k\Pe},\\
  D_{\rm eff}^{\rm full}(k)
  &=
  \frac{\operatorname{Re}\alpha(k)}{k^2},\\
  \kappa_N^{\rm full}(k)
  &=
  \frac{D_{\rm eff}^{\rm full}(k)-K_{{\rm dir},N}}{\Pe^2}.
\end{align}
\end{subequations}
The reported values are least-squares fits with no intercept: the imaginary part is fit as an odd term proportional to \(k\), and the real part as an even term proportional to \(k^2\). Dense generalized eigensolves are used on the default validation meshes; a sparse shift-invert option targets the same branch for larger meshes.

\begin{figure}
  \centering
  \includegraphics[width=0.92\linewidth]{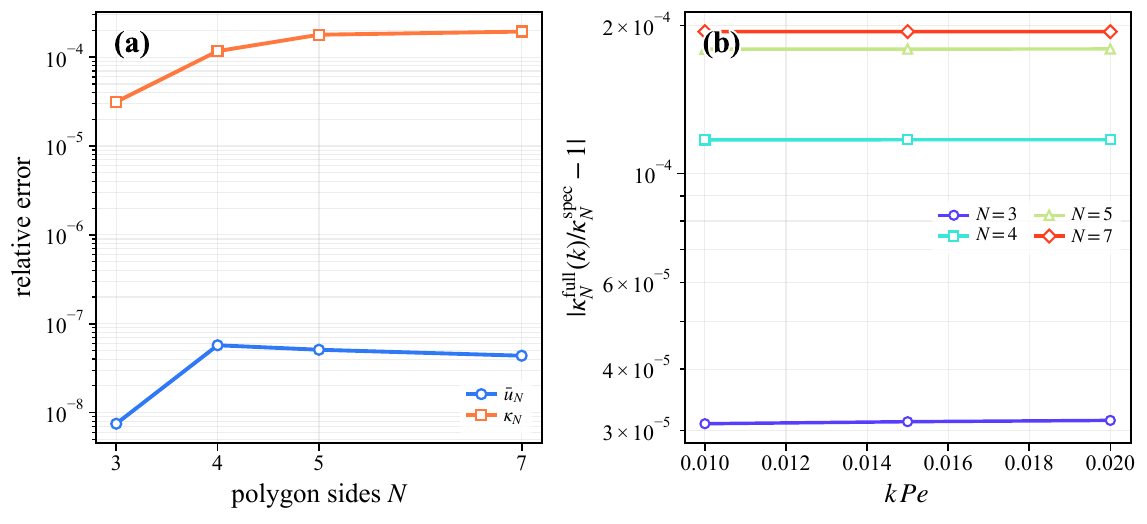}
  \caption{Full transverse-space Fourier--FEM small-wavenumber validation for the high-shear reduced spectral cases. Panel (a) compares the fitted full-space \(\bar u_N\) and \(\kappa_N\) with the 64-mode spectral values. Panel (b) shows the per-wavenumber \(\kappa_N^{\rm full}(k)\) errors before the least-squares fit.}
  \label{fig:app-full-fourier-validation}
\end{figure}

We apply this check to a high-shear validation parameter set, \(p=100\), \(\Pe=10^4\), \(\Per=10^3\) and \(N=3,4,5,7\). The full transverse-space Fourier--FEM calculation uses validation meshes with \(n_r=12\) and \(n_\theta=6N\), while the reduced model keeps 64 transverse modes on the same meshes. The small-wavenumber samples are \(k=10^{-6}\), \(1.5\times10^{-6}\) and \(2\times10^{-6}\), so that \(k\Pe\le0.02\). In Table~\ref{tab:app-full-fourier-validation} and Figure~\ref{fig:app-full-fourier-validation}, the superscript ``spec'' denotes the 64-mode reduced spectral value and ``full'' denotes the value extracted from \(\alpha(k)\).

\begin{table}
  \centering
  \caption{Full transverse-space Fourier--FEM validation of the reduced spectral model for the high-shear validation cases. The errors are \(\epsilon_{\bar u}=|\bar u_N^{\rm full}/\bar u_N^{\rm spec}-1|\) and \(\epsilon_\kappa=|\kappa_N^{\rm full}/\kappa_N^{\rm spec}-1|\).}
  \begin{tabular}{cccccc}
\toprule
$N$ & nodes & $\lambda_1$ & $\bar u_N^{\rm spec}$ & $\epsilon_{\bar u}$ & $\epsilon_\kappa$ \\
\midrule
3 & 217 & 1.3103 & 0.438319 & $7.50\times 10^{-9}$ & $3.14\times 10^{-5}$ \\
4 & 289 & 2.1737 & 0.469039 & $5.73\times 10^{-8}$ & $1.17\times 10^{-4}$ \\
5 & 361 & 2.4939 & 0.480929 & $5.10\times 10^{-8}$ & $1.79\times 10^{-4}$ \\
7 & 505 & 2.7311 & 0.489622 & $4.37\times 10^{-8}$ & $1.94\times 10^{-4}$ \\
\bottomrule
\end{tabular}

  \label{tab:app-full-fourier-validation}
\end{table}

The Fourier--FEM check in \eqref{eq:app-full-fourier-operator} keeps the complete finite-element transverse space for the leading advection--relaxation operator rather than using a modal truncation. Its small-\(k\) expansion tests the sampling speed \(\bar u_N\), the leading Taylor coefficient \(\kappa_N\), and the direct axial diffusivity \(K_{{\rm dir},N}\) when \(M_B\) is retained. The \(A\)-dependent Fourier couplings associated with \(-\e_sA c_z\) and \(-\partial_s(Ac)\) are lower-order terms in the high-\(\Pe\), long-wave ordering of Section~\ref{sec:taylor-aris-reduction}; the conservative axial contribution is the drift \(U_{A,N}\) derived in Section~\ref{subsec:direct-cross-diffusion}.

\end{document}